\documentclass[fleqn,usenatbib]{mnras}

\usepackage{newtxtext,newtxmath}

\usepackage[T1]{fontenc}
\usepackage{placeins}

\DeclareRobustCommand{\VAN}[3]{#2}
\let\VANthebibliography\thebibliography
\def\thebibliography{\DeclareRobustCommand{\VAN}[3]{##3}\VANthebibliography}

\usepackage{graphicx}	
\usepackage{amsmath}	
\usepackage{float}
\usepackage{caption}    
\usepackage[switch]{lineno}
\usepackage{multicol}
\usepackage{fix-cm}
\title[MMDCS: Diffuse Radio Emission in High-$z$ Clusters]{The MeerKAT Massive Distant Clusters Survey: detection of diffuse radio emission in galaxy clusters at $z > 1$}

\author[D. G. Phuravhathu et al.]{Dakalo G. Phuravhathu$^{1}$\thanks{E-mail: dakalophuravha2@gmail.com},
M. Hilton$^{1,2}$,
S. P. Sikhosana$^{2,3}$,
Y. C. Perrott$^{4}$,
T. Mroczkowski$^{5}$,
L. Di Mascolo$^{6,7}$,
\newauthor
D. Y. Klutse$^{2,3}$,
K. Knowles$^{3,8,9}$,
J. van Marrewĳk$^{10}$,
K. Moodley$^{2,3}$,
B. Partridge$^{11}$,
C. Sifón$^{12}$,
U. Sureshkumar$^{1}$,
\newauthor
E. J. Wollack$^{13}$
\\
$^{1}$Wits Centre for Astrophysics, School of Physics, University of the Witwatersrand, Private Bag 3, 2050, Johannesburg, South Africa\\
$^{2}$School of Mathematics, Statistics \& Computer Science, University of KwaZulu-Natal, Westville Campus, Durban 4041, South Africa\\
$^{3}$Astrophysics Research Centre, University of KwaZulu-Natal, Durban, 3696, South Africa\\
$^{4}$School of Chemical and Physical Sciences, Victoria University of Wellington, PO Box 600, Wellington 6140, New Zealand\\
$^{5}$European Southern Observatory, Karl-Schwarzschild-Str. 2, 85748 Garching, Germany\\
$^{6}$Kapteyn Astronomical Institute, University of Groningen, Landleven 12, 9747 AD Groningen, The Netherlands\\
$^{7}$Université Côte d’Azur, Observatoire de la Côte d’Azur, CNRS, Laboratoire Lagrange, France\\
$^{8}$Centre for Radio Astronomy Techniques and Technologies, Department of Physics and Electronics, Rhodes University, P.O. Box 94, Makhanda 6140, South Africa\\
$^{9}$South African Radio Astronomy Observatory, 2 Fir Street, Black River Park, Observatory, Cape Town 7925, South Africa\\
$^{10}$Leiden Observatory, Leiden University P.O. Box 9513, 2300 RA Leiden The Netherlands\\
$^{11}$Department of Astronomy, Haverford College, Haverford, PA 19041, USA\\
$^{12}$Instituto de Física, Pontificia Universidad Católica de Valparaíso, Casilla 4059, Valparaíso, Chile\\
$^{13}$NASA Goddard Space Flight Center, 8800 Greenbelt Rd, Greenbelt, MD 20771, USA
}

\date{Accepted XXX. Received YYY; in original form ZZZ}

\pubyear{2025}

\begin{document}
\label{firstpage}
\pagerange{\pageref{firstpage}--\pageref{lastpage}}
\maketitle

\begin{abstract}
Diffuse, low surface-brightness radio emission in merging galaxy clusters provides insights into cosmic structure formation, the growth of magnetic fields, and turbulence. This paper reports a search for diffuse radio emission in a pilot sample of six high-redshift ($1.01 < z < 1.31$) galaxy clusters from the MeerKAT Massive Distant Cluster Survey (MMDCS). These six clusters are selected as the most massive ($M_{\rm 500c} = 6.7\,- 8.5 \times 10^{14}~\rm{M_{\odot}}$) systems based on their Sunyaev–Zel’dovich mass from the full MMDCS sample of 30 ACT DR5 clusters, and were observed first to explore the high-mass, high-redshift regime. Diffuse radio emission is confidently detected in four clusters and tentatively identified in two, with $k$-corrected radio powers scaled to 1.4 GHz ranging from $(0.30 \pm 0.08)$ to $(3.55 \pm 1.06) \times 10^{25} \, \mathrm{W\,Hz^{-1}}$ and linear sizes between 0.47 and 1.08 Mpc. Combining \textit{Chandra} X-ray data with MeerKAT radio data, we find that 80$\%$ of clusters with X-ray observations exhibit disturbed morphologies indicative of mergers. These $z > 1$ galaxy clusters scatter around the established radio power-mass scaling relation observed at lower redshifts, supporting turbulent re-acceleration models in high-redshift mergers. However, their radio spectra are predicted to steepen ($\alpha < -1.5$) due to enhanced inverse Compton losses in the cosmic microwave background, rendering them under-luminous at 1.4 GHz and placing them below the correlation. Our results demonstrate that merger-driven turbulence can sustain radio halos even at $z > 1$  while highlighting MeerKAT’s unique ability to probe non-thermal processes in the early universe. 
\end{abstract}




\section{Introduction}

Galaxy clusters, the largest gravitationally bound structures in the universe, form at the intersections of cosmic web filaments through a hierarchical process involving gravitational in-fall and mergers with smaller sub-clusters. These merger events, among the most energetic phenomena since the Big Bang, release tremendous energy ($\sim$10$^{64}\, \mathrm{erg}$), a significant portion of which dissipates as shocks and turbulence within the intra-cluster medium  \citep[ICM,][]{2007PhR...443....1M,2002ASSL..272....1S,2003PhPl...10.1992S}. The turbulent activity heats the ICM to temperatures of $10^{7}$--$10^{8}\, \text{K}$, amplifies magnetic fields, and (re)accelerates cosmic ray electrons (CRe) to relativistic energies. These processes result in the emission of diffuse synchrotron radiation, observed as radio halos and relics, which stem from the ICM rather than individual galaxies. However, it is important to note that while this mechanism is generally preferred, alternative models, such as hadronic processes, have been proposed and are still actively debated \citep[see review by][]{2014IJMPD..2330007B}.
\newline

Radio halos and relics are large-scale radio sources ($\sim$$0.1$--$1\, \text{Mpc}$) that emit steep-spectrum radiation ($-1.0 \gtrsim \alpha \gtrsim -1.4$\footnote[1]{$S_{\nu} \propto \nu^{\alpha},$ where $S_{\nu}$ is the source flux density at frequency $\nu$ and $\alpha$ is the spectral index}) with very low surface brightness \citep[$\sim$$\mu$Jy/arcsec$^{2}$,][]{2013A&A...551A.141M,2019SSRv..215...16V,2020MNRAS.499..404P}. However, they differ in terms of their location, morphology, and polarisation properties. Radio halos are centrally located within clusters and often trace the X-ray emitting ICM, with a regular morphology and low polarisation. The formation of radio halos is primarily attributed to the re-acceleration of pre-existing electrons due to turbulence within the ICM during cluster mergers, with the re-acceleration process facilitated by magnetohydrodynamic (MHD) turbulence \citep{2001MNRAS.320..365B,2005MNRAS.357.1313C}. Radio halos have been found predominantly in merging clusters and show a close correlation with the X-ray emission from the ICM, further supporting the idea that their origin is tied to the thermal and non-thermal components of the cluster's central ICM \citep{2014IJMPD..2330007B,2010ApJ...721L..82C}. 
\newline

In contrast, radio relics are located at the peripheral regions of galaxy clusters, often appearing as elongated, arc-shaped structures with high polarisation (up to 70$\%$). Their formation mechanism is thought to originate from shock waves propagating through the ICM during cluster mergers, which accelerate electrons to relativistic speeds via the diffusive shock acceleration \citep[DSA,][]{1998A&A...332..395E}. The shock fronts, which typically have a Mach number between 1 and 3, are not always strong enough to accelerate electrons from the thermal pool alone \citep{2017NatAs...1E...5V,2024ApJ...962..161S}, and thus, pre-accelerated populations of electrons are considered essential for the formation of relics \citep{,2007MNRAS.375...77H,2023A&A...675A..51D}. Observational evidence suggests that relics often occur in merging clusters with non-relaxed dynamics. Their morphology can sometimes resemble that of shock waves, supporting the theory that they trace the outer edges of cluster mergers \citep{2016MNRAS.460L..84B,2020A&A...636A..30R}.
\newline

The morphology of both the X-ray emission from the ICM \citep{1988xrec.book.....S,2000ApJS..129..435B} and the distribution of galaxies within clusters provide valuable insights into their dynamical state \citep{1980ApJ...236..351D}. Galaxy clusters can be broadly classified as either dynamically relaxed or disturbed, with the latter often indicative of ongoing or recent merger activity \citep{2012MNRAS.420.2120M,2020MNRAS.497.5485Y,2022MNRAS.513.3013Y}. Dynamically disturbed clusters show irregular morphologies and substructure, and often exhibit multiple brightest cluster galaxies (BCGs), while relaxed clusters tend to have a smooth, symmetric X-ray appearance centred on a single BCG \citep{2015A&A...575A.127P,2020MNRAS.497.5485Y}. However, mergers that occur along the line of sight are often difficult to identify in projected X-ray images because they may not produce clear asymmetries or substructure. Mergers play a crucial role in generating the turbulence necessary for (re)accelerating cosmic ray electrons and producing diffuse radio emission \citep{2014IJMPD..2330007B,2019SSRv..215...16V}. Specifically, cluster mergers drive shocks and turbulence within the ICM that can amplify magnetic fields \citep{2002ARA&A..40..319C} and accelerate particles \citep{2008SSRv..134..207P}. Therefore, understanding the dynamical state of galaxy clusters is essential for interpreting the origin and properties of observed radio halos and relics. In this paper, we will consider the X-ray morphology (using archival \textit{Chandra} X-ray data) of our cluster sample together with the detected diffuse radio emission to gain insights into the dynamical processes at play in these high-redshift systems, contributing to the broader understanding of the link between cluster mergers and radio halo formation.
\newline

Diffuse radio emission is difficult to observe at higher redshifts ($z > 1$) due to three key reasons: the increased energy density of the  cosmic microwave background (CMB) leads to more rapid energy losses of cosmic ray electrons through inverse Compton (IC) scattering; redshift dimming reduces the surface brightness of the emission; and magnetic fields may not yet be strong enough to sustain bright radio halos, further limiting synchrotron luminosity \citep{2014IJMPD..2330007B,2021NatAs...5..268D}. These effects are predicted to suppress the occurrence of diffuse radio emission in high-redshift clusters. At lower redshifts $(0.22<z<0.65)$, \citet{2021MNRAS.504.1749K} report a detection rate of $\sim\!70\%$,  while \citet{2021NatAs...5..268D, 2025A&A...695A.215D} report a decline in detection rates from $\sim\!50\%$ at $0.6 < z < 0.9$ to $\sim\!9\%$ at $0.78 \le z \le 1.5$. This redshift-dependent suppression aligns with the $(1+z)^4$ scaling of IC losses and lower cluster masses of high-$z$ clusters \citep{2014IJMPD..2330007B}. Observational challenges have limited the number of known high-redshift halos and relics, with only a few detections reported beyond $z > 0.8$. Notable examples include the radio halo in "El Gordo" \citep[ACT-CL J0102-4915,][]{{2014ApJ...786...49L}} at $z = 0.870$ and the $z = 1.23$ halo in ACT-CL J0329.2-2330 \citep{2025A&A...698L..17S}, with the latter being a part of our sample. Recent studies continue to push these limits, with the detection of diffuse radio emission at $z > 1.3$ \citep{2025A&A...695A.215D} and a candidate mini-halo at $z = 1.71$ reported by \citet{2025ApJ...987L..40H}.
\newline

Recent advancements in observational technology have allowed more effective studies of diffuse radio emission using next-generation Square Kilometre Array (SKA) precursor telescopes such as MeerKAT \citep{2016mks..confE...1J,2018ApJ...856..180C}, LOFAR \citep{2013A&A...556A...2V}, and ASKAP \citep{2007PASA...24..174J}. MeerKAT stands out as the most sensitive instrument currently operational in the Southern Hemisphere, comprising a total of 64 dishes situated in South Africa's Karoo region \citep{2016mks..confE...1J,2018ApJ...856..180C,2020ApJ...888...61M}. Its wide frequency coverage ($UHF$: 580–1015 MHz; $L$: 900–1670 MHz; $S$: 1750–3500 MHz) enables spectral studies of diffuse emission both within individual bands (in-band studies) due to their wide bandwidths and across multiple bands for broader frequency coverage. Approximately three-quarters of its antennas are concentrated within a dense core spanning a radius of just 1 km, enhancing sensitivity to extended structures while its outer antennas provide sufficient resolution for compact sources even at higher redshifts. This capability allows MeerKAT to effectively observe low surface brightness emission while resolving compact sources, thereby facilitating significant advancements in our understanding of diffuse emissions within galaxy clusters.
\newline

In this paper, we present the findings of a search for diffuse emission in an initial sub-sample of six clusters selected from the MeerKAT Massive Distant Clusters Survey. In Section \ref{sec:Observations and data reduction}, we describe the sample selection and outline our MeerKAT observations and data reduction. The  MeerKAT results are presented in Section \ref{sec:Results}, followed by a discussion in Section \ref{sec:Discussion}, and a summary and conclusions in Section \ref{sec:SUMMARY AND CONCLUSIONS}. Throughout this paper, $R_{\rm 500c}$ is defined as the radius within which the mean mass density of the cluster is 500 times the critical density of the universe at the cluster’s redshift. $M_{\rm 500c}$ corresponds to the total mass (dark matter, gas, and galaxies) enclosed within $R_{\rm 500c}$. We adopt the standard $\Lambda$CDM cosmology with $H_{0} = \mathrm{70\, km\, s^{-1}\, Mpc^{-1}}$, $\Omega_{\rm{m}} = 0.3$ and $\Omega_{\Lambda} = 0.7$. 
\newline

\section{Observations and data reduction}
\label{sec:Observations and data reduction}
\subsection{The cluster sample}
\begin{figure}
 \includegraphics[width=\columnwidth]{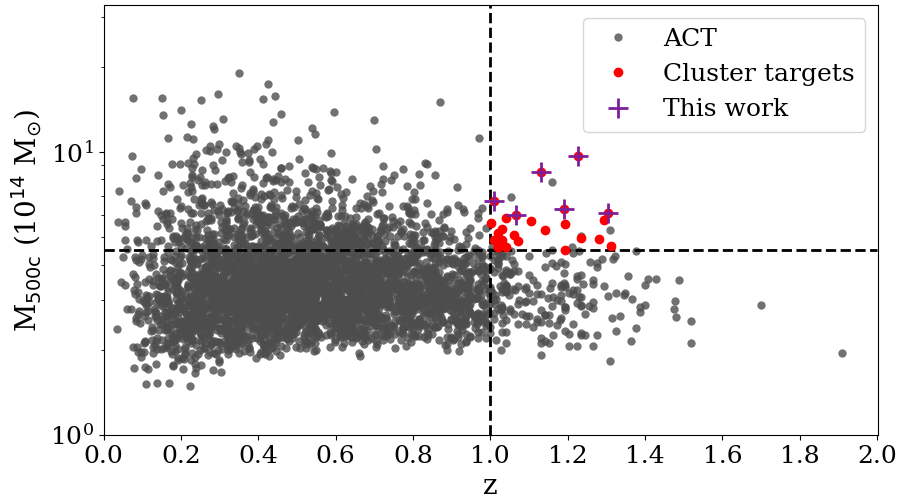}
 \caption{Mass versus redshift plot showing the ACT SZ-selected galaxy cluster  catalogues (gray dots). The 30 MeerKAT target sample (red dots) are ACT clusters above $z > 1$ with mass ($M_{\rm 500c}$) $> 4.5\, \times\, 10^{14}$\,M$_{\odot}$, and within the DECaLS DR9 optical+IR survey (the dashed lines indicate the mass and $z$ limits used; the 6 objects in the mass--$z$
selection range that are not targeted are outside the DECaLS DR9 survey footprint). The purple crosses overlaid on six of these red dots indicate the specific clusters selected for this study.}
\label{fig:1}
\end{figure}

The MeerKAT Massive Distant Cluster Survey (MMDCS) is a MeerKAT $L/UHF$-band survey specifically designed to study a selection of 30 massive galaxy clusters at $z > 1$. The survey aims to study various aspects of these high-redshift clusters, including radio emission, star formation, active galactic nucleus (AGN) activity, and environmental effects within the cluster environments. The MMDCS sample comprises 30 of the most massive clusters selected from the Atacama Cosmology Telescope \citep[ACT;][]{2011ApJS..194...41S} Data Release 5 catalogue \citep{2021ApJS..253....3H}. For this pilot study, we observed the six most massive clusters in the MMDCS sample (see Figure \ref{fig:1}). These clusters were prioritised to maximise sensitivity to diffuse emission in the high-mass regime, given the expected correlation between radio halo power and cluster mass \citep{2013ApJ...777..141C}. 
\newline
\newline
This  catalogue contains over 4000 clusters, with 222 located at $z > 1$. The ACT  catalogue is based on the Sunyaev-Zel’dovich (SZ) effect \citep{1972CoASP...4..173S}, which is a powerful method for detecting galaxy clusters due to several unique properties: its redshift-independent surface brightness, the distinct spectral distortion of the CMB with a negative signal (decrement) at frequencies below 218 GHz, and the integrated flux density tracing thermal energy, which serves as a reliable proxy for total cluster mass. We restrict the selection to clusters with $z > 1$ and $M_{\rm 500cCal}$\footnote[2]{$M_{\rm 500cCal}$ refers to the richness-based, weak-lensing-calibrated mass estimate, which is considered the most accurate mass proxy available in ACT DR5 \citep{2021ApJS..253....3H}.} $> 4.5\, \times\, 10^{14}\, \rm{M_\odot}$, and which fall within the Dark Energy Camera Legacy Survey (DECaLS) DR9 optical+IR survey footprint \citep{2019AJ....157..168D}, which provides essential data for source redshifts and classification (see Figures \ref{fig:7}-\ref{fig:9}). The target cluster sample covers redshift range $1.00 < z < 1.31$ and mass range $4.5 < M_{\rm 500c} / 10^{14}\,\rm{M_{\odot}} < 9.7$ (see Figure \ref{fig:1}). The properties of the ACT SZ clusters properties are summarised in Table \ref{table:1}. In this study, we report on MeerKAT observations of a sub-sample comprising the six most massive clusters from the MMDCS sample. The full MMDCS sample will be analysed in future work.

\begin{table*}
\caption{Properties of of the first 6 clusters in the sample.}
\label{table:1}
\begin{tabular}{cccccccc}
\hline
Cluster Name    & RA (J2000)   & Dec (J2000)  & redshift             & \multicolumn{1}{l}{redshift type} & \textit{$M_{\rm 500cCal,SZ}$} & $M_{\rm 500cUncorr,SZ}$ & SNR \\
(ACT-CL) & (hh:mm:ss.s) & (dd:mm:ss.s) & \textit{z}           & \multicolumn{1}{l}{}              & \multicolumn{2}{c}{($10^{14} M_{\odot}$)}               &                    \\ \hline
J1137.8+0728    & 11:37:50.8   & +07:28:36.0  & 1.01 $\, \pm\, 0.03$ & phot$^{\mathrm{a}}$                             & 6.7$^{+1.2}_{-1.1}$           & 5.3$^{+0.9}_{-0.7}$     & 16.9                  \\
J0546.6$-$5345    & 05:46:37.6   & $-$53:45:33.2  & 1.07                 & spec$^{\mathrm{b}}$                              & 6.0$^{+1.1}_{-1.0}$           & 4.7$^{+0.8}_{-0.7}$     & 12.8                  \\
J2106.0$-$5844    & 21:06:04.3   & $-$58:44:35.0  & 1.13                 & spec$^{\mathrm{c}}$                              & 8.5$^{+1.5}_{-1.4}$           & 6.8$^{+1.1}_{-0.9}$     & 18.7                  \\
J1142.7+1527    & 11:42:46.0   & +15:27:22.0  & 1.19                 & spec$^{\mathrm{d}}$                              & 6.3$^{+1.1}_{-1.0}$           & 5.0$^{+0.8}_{-0.7}$     & 19.6                  \\
J0329.2$-$2330    & 03:29:17.7   & $-$23:30:09.7  & 1.23                 & spec$^{\mathrm{c}}$                              & 9.7$^{+1.7}_{-1.6}$           & 8.0$^{+1.3}_{-1.1}$     & 22.1                  \\
J0003.9+1642    & 00:03:51.8   & +16:42:06.5  & 1.31$\, \pm\, 0.03$  & phot$^{\mathrm{a}}$                              & 6.1$^{+1.1}_{-1.0}$           & 4.9$^{+0.8}_{-0.7}$     & 14.7                  \\ \hline
\end{tabular}
\begin{minipage}{\textwidth}
\footnotesize
\textbf{Notes:} The SZ values are obtained from the ACT DR5 catalogue \citep{2021ApJS..253....3H}. Columns: (1) ACT DR5 cluster name; (2) J2000 Right Ascension of the SZ centre; (3) J2000 Declination of the SZ centre; (4-5) redshift and redshift type of the cluster, respectively. The redshifts are sourced from: $^{\mathrm{a}}$zCluster \citep{2021ApJS..253....3H}, $^{\mathrm{b}}$ACT \citep{2013A&A...551A.141M,2016MNRAS.461..248S,2018ApJS..235...20H}, $^{\mathrm{c}}$South Pole Telescope (SPT) \citep{2019ApJ...878...55B,2020ApJS..247...25B}, $^{\mathrm{d}}$Massive and Distant Clusters of WISE Survey (MaDCoWS) \citep{2015ApJ...812L..40G}; (6-7) ACT SZ mass calibrated using weak-lensing and uncorrected ACT SZ mass, respectively. Note that the ACT SZ uncorrected mass is not corrected for Eddington bias and is used for comparison with Planck PSZ2 masses; (8) Signal-to-Noise ratio from ACT.
\end{minipage}
\end{table*}

\subsection{Observation}
The observations were conducted using the MeerKAT $L$-band receiver (PI: M. Hilton and proposal ID: SCI-20220822-MH-01), which operates between 856 and 1712 MHz, with a native bandwidth of 856 MHz and a central frequency of 1.28 GHz. The data were recorded using 4096 frequency channels and an integration time of 8 seconds per visibility, resulting in a total on-target time of 3.5 hours per cluster, except for ACT-CL J2106.0–5844, which was observed for approximately 6 hours. Primary calibrators were observed for 10 minutes at the start of each session for band-pass and flux calibration. Secondary calibrators were used for complex-gain and delay calibration, alternating with target observations every 2 and 15 minutes, respectively. The details of the MeerKAT observations are summarised in Table \ref{table:2}.

\subsection{Data Reduction and imaging}

We reduced the data using the \texttt{Oxkat} v0.4\footnote[3]{\url{https://github.com/IanHeywood/oxkat}} software \citep{2020ascl.soft09003H}, a semi-automated tool designed for processing MeerKAT data. The \texttt{Oxkat} pipeline, consisting of Python-based scripts, streamlines the data reduction process by incorporating various radio astronomy software packages, such as \texttt{CASA} v5.6\footnote[4]
{\url{https://casa.nrao.edu}} 
\citep{2007ASPC..376..127M,2022PASP..134k4501C} and \texttt{WSClean v3}\footnote[5]
{\url{https://sourceforge.net/p/wsclean/wiki/Home/}} \citep{2014MNRAS.444..606O}. This pipeline handles first-generation calibration (1GC: cross-calibration and initial imaging), flagging, second-generation calibration (2GC: self-calibration and imaging), and third-generation calibration (3GC: direction-dependent calibration), following a structured approach that includes several key steps outlined below.
\newline
\newline
The first step involves  initial calibration and flagging. The raw Measurement Set (MS) is 
duplicated and averaged over 1024 channels to optimize continuum data processing and 
reduce computation time. Primary calibration begins by re-phasing the primary calibrator 
using the CASA task  \texttt{fixvis}, followed by Radio Frequency Interference (RFI) flagging, where known RFI channels and contaminated bands (up to $\sim 40\%$ of the original bandwidth) are flagged using the \texttt{CASA} \texttt{flagdata} task and auto flaggers. Auto flagging tools, like \texttt{tfcrop} and \texttt{rflag} in \texttt{CASA}, along with the  \texttt{tricolour} v0.1.8\footnote[6]{\url{https://github.com/ratt-ru/tricolour}} package \citep{2022ASPC..532..541H}, are employed to further clean the data of any residual RFI. The Stevens-Reynolds 2016 flux density scale \citep{2016ApJ...821...61P} is applied to the primary calibrator. Secondary calibration follows, constructing an intrinsic model for the secondary calibrator based on the primary calibrator data. Delay, band-pass, and gain calibrations are iteratively derived from both calibrators, and these solutions are applied to all calibrators and target data. These solutions are applied to the target data, ensuring calibration integrity through multiple rounds of residual flagging.
\newline
\newline
Following cross calibration, the target data are imaged using \texttt{WSClean}. A \texttt{Briggs} weighting \citep{1995AAS...18711202B} of $-0.3$ is used to balance noise sensitivity and angular resolution. \texttt{WSClean} uses multi-scale and wide-band deconvolution algorithms to improve the imaging of diffuse emission, deconvolving eight sub-band images with central frequencies of 909, 1016, 1123, 1230, 1337, 1444, 1551, and 1658 MHz, and generating a multi-frequency synthesis (MFS) map. Self-calibration is performed using the \texttt{Cubical} software \citep{2018MNRAS.478.2399K}, which refines calibration solutions through iterative re-imaging and masking, significantly improving the images' dynamic range.
\newline
\newline
For datasets where 2GC falls short in correcting direction-independent effects, resulting in images with significant residual artifacts, we employ direction-dependent calibration. The 3GC calibration step addresses direction-dependent effects (DDEs) in the Radio Interferometer Measurement Equation (RIME)  
\citep{2010A&A...524A..61N,2011A&A...527A.106S,2011A&A...527A.107S,2011A&A...527A.108S,2011A&A...531A.159S,2015MNRAS.449.2668S}. For MeerKAT, major DDEs include primary beam rotation 
and pointing errors, which manifest as radial artifacts around bright sources. For wide-field imaging, phase and amplitude gains may vary significantly across the field of view, necessitating direction-dependent calibration. The 
\texttt{Oxkat} pipeline employs a facet-based approach using the \texttt{killMS}\footnote[7]{\url{https://github.com/saopicc/killMS}} \citep[kMS,][]{2014arXiv1410.8706T,2014A&A...566A.127T} and \texttt{DDFacet}\footnote[8]{\url{https://github.com/saopicc/DDFacet}} \citep[DDF,][]{2018A&A...611A..87T} software to account for DDEs. The kMS tool 
performs 3GC by exploiting Wirtinger complex differentiation to directly estimate the 
physical DD RIME terms per facet, while DDF is a facet-based imager that can take into account kMS 
solutions to perform wide-band, wide-field differential spectral deconvolution.

\begin{table*}
\caption{Summary of MeerKAT $L$-band observations for the cluster sample, including observational parameters and both full-resolution (FR) and low-resolution (LR) imaging properties.}
\label{table:2}
\begin{tabular}{ccccccccc}
\hline
Cluster Name                & Observing date & Bandpass   & Phase & Flagged      & $\mathrm{\theta_{synth, FR}}$         & $\mathrm{\sigma_{rms, FR}}$ & $\mathrm{\theta_{synth, LR}}$         & $\mathrm{\sigma_{rms, LR}}$ \\
(ACT-CL)             & (Y-M-D)      & calibrator & calibrator & ($\%$) & ($\mathrm{'' \times ''}$, $^{\circ}$) & $\mathrm{\mu Jy/beam}$      & ($\mathrm{'' \times ''}$, $^{\circ}$) & $\mathrm{\mu Jy/beam}$      \\ \hline
J1137.8$+$0728$^{\textit{*}}$                & 2023-02-01     & J0408$-$6545 & J1150$-$0023 & 43.3 & 8.6 $\times$ 5.8, 174.1                 & 4.8                         & 15.5 $\times$ 15.5, 0.0               & 7.5                         \\
J0546.6-5345                & 2023-02-04     & J0408$-$6545 & J0538$-$4405 & 47.9 & 5.6 $\times$ 5.6, 0.0                 & 6.1                         & 15.2 $\times$ 15.2, 0.0               & 9.4                         \\
J2106.0-5844$^{\textit{*}}$ & 2022-12-19     & J1939$-$6342 & J1939$-$6342 & 52.3 & 9.0 $\times$ 5.3, 56.4                & 4.5                         & 14.9 $\times$ 14.3, 130.3             & 6.5                         \\
J1142.7$+$1527                & 2023-02-04     & J0408$-$6545 & J1120+1420 & 43.2 & 7.9 $\times$ 7.9, 0.0                 & 5.4                         & 15.6 $\times$ 15.6, 0.0               & 7.2                         \\
J0329.2-2330$^{\textit{*}}$ & 2023-01-27     & J0408$-$6545 & J0409$-$1757 & 48.1 & 6.5 $\times$ 6.5, 0.0                 & 5.0                         & 15.2 $\times$ 15.2, 0.0               & 6.1                         \\
J0003.9+1642$^{\textit{*}}$ & 2023-01-29     & J1939$-$6342 & J2253+1608 & 44.7 & 10.2 $\times$ 5.5, 167.5              & 7.4                         & 16.2 $\times$ 14.9, 175.1             & 9.2                         \\ \hline
\end{tabular}
\begin{minipage}{\textwidth}
\footnotesize
\textbf{Notes:} Columns: (1) ACT DR5 cluster name; (2) Observation date; (3-4) Bandpass and phase calibrators; (5) The percentage of MeerKAT data that is flagged during reduction, which includes frequency ranges that are known to be affected by satellite interference; (6-9) synthesized beam parameters (major and minor axes, and position angle) and central rms noise levels for both full and low-resolution images.$^{\textit{*}}$Direction-dependent corrections were required for the cluster field. The image shows significant variability in noise, attributed to residual contamination from the artefacts of the bright source.
\end{minipage}

\end{table*}

\subsection{Compact source subtraction}

We performed the subtraction of compact sources to ensure that the flux density measurements of the diffuse emission remained uncontaminated by any compact sources emissions. This process was carried out in the $uv$-plane, targeting all sources within the cluster radius ($R_{500c}$), centred at the cluster’s Sunyaev-Zel’dovich (SZ) centre, using \texttt{CRYSTALBALL} v0.4.1\footnote[9]{\url{https://github.com/caracal-pipeline/crystalball}} and \texttt{MSUTILS} v1.2.0\footnote[10]
{\url{https://github.com/SpheMakh/msutils}}. \texttt{CRYSTALBALL} uses a source list generated by \texttt{WSClean} to create a model column in the MS file, containing only the tagged compact sources. The source list was generated from an image produced by \texttt{WSClean} using a taper-inner-tukey of $10\, \rm{k\lambda}$ and Briggs weighting of 0 (see the left panel of Figures \ref{fig:10}-\ref{fig:15}), which effectively suppresses diffuse emission and isolates compact sources. \texttt{MSUTILS} then subtracts this model column from the calibrated data column to create a new column of compact-source-subtracted visibilities. These visibilities are then used to produce low-resolution maps, enhancing the diffuse emission and ensuring no blending occurs between compact and extended emission. 

\subsection{Low-resolution maps}
To enhance the detection of large-scale diffuse emission, we created low-resolution images by applying an outer taper of $\leq 12\, \rm{k}\lambda$, which corresponds to an angular resolution of $\sim17.2^{\prime\prime}$, or a physical scale of $\sim139.5\, \rm{kpc}$ at the cluster redshift of $z = 1.01$, and using Briggs weighting with a robust parameter of 0 (see the right panel of Figures \ref{fig:10}-\ref{fig:15}). This approach gives more weight to shorter baselines, which are more sensitive to extended emission. The integrated flux densities were obtained from these low-resolution images using polygon regions around each source, tracing the $3\mathrm{\sigma_{rms}}$ radio contours to comprehensively encompass the extent of the diffuse radio emission. The uncertainties of the flux densities of the diffuse radio sources are estimated using the expression

\begin{equation}
    \Delta{S_{\nu}} = \sqrt{(\delta{S_{\nu}} \times S_{\nu})^{2} + N_{\mathrm{beams}}\sigma^{2}_{\mathrm{rms}} + \sigma^{2}_{\mathrm{sub}}},
\end{equation}
where $\delta{S_{\nu}}$ represents the uncertainty in the absolute flux calibration, assumed to be 5$\%$ for $L$-band, $N_{\mathrm{beams}}$ denotes the number of beams within the region where the flux density was measured, and $\sigma_{\mathrm{rms}}$ denotes the local rms noise of the image. The uncertainty attributable to compact-source subtraction, $\sigma_{\mathrm{sub}}$, is given by

\begin{equation}
    \sigma^{2}_{\mathrm{sub}} = \sum_{i} N_{\mathrm{beams,i}} \sigma^{2}_{\mathrm{rms}},
\end{equation}
where the summation is taken over all the $i$ sources that were subtracted within the polygon region  surrounding the diffuse radio source, with $N_{\mathrm{beams,i}}$ representing the number of beams covering the $i^{th}$ compact source. 
\newline
\newline
The $k$-corrected radio power was calculated using the expression 

\begin{equation}
   P_{1.4 \text{ GHz}} = 4\pi D_L^2 \frac{S_{1.4 \text{ GHz}}}{(1 + z)^{\alpha+1}}
\end{equation}
where $S_{\rm 1.4\,GHz}$ is the observed ﬂux density scaled to 1.4 GHz, $D\mathrm{_{L}}$ is the luminosity distance, and $\alpha$ is the spectral index. While we attempted to produce in-band spectral index maps with $L$-band observations, the faintness of the emission made this challenging, and reliable maps could not be generated. Instead, we assume a Gaussian distribution for the spectral index $\alpha$ with a mean of $-1.3$ and a standard deviation of $0.4$ for calculating the radio luminosity \citep{2021MNRAS.504.1749K}. This assumption is based on the typical range of spectral indices observed in radio halos, which often cluster around $-1.3$ but can vary. By using a Gaussian distribution, we account for the uncertainty in spectral indices and the possibility that they may be steeper than those found in local systems, which is consistent with observations of high-redshift clusters where the environment may influence the spectral properties.

\section{Results}
\label{sec:Results}
In this section, we present the findings from the analysis of six galaxy clusters, focusing on their diffuse emission. Four of the clusters contain prominent diffuse emission that is visible in the full-resolution (FR) images (see Figure \ref{fig:2a}). In the remaining two clusters, the diffuse emission is undetectable in full-resolution images but becomes apparent only after compact source subtracted low-resolution (LR) imaging (see Figure \ref{fig:2b}). The detections meet the $3\sigma$ significance threshold, confirming the presence of diffuse emission. Below, we provide an overview of these results and highlight key features of the diffuse emission across the galaxy clusters. The measured properties are provided in Table \ref{table:3}.

\subsection{Prominent Diffuse Emission}
\label{sec:Prominent Diffuse Emission}

\subsubsection{ACT-CL J2106.0-5844}

ACT-CL J2106.0-5844 (hereafter, ACT-CL J2106)\footnote[11]{J2106 will be investigated in more detail in Perrott et al (in prep).}, also known as SPT-CL J2106-5844, is among the most massive galaxy clusters identified at a high redshift of {\it z} = 1.13, with a mass of $M_{\rm 500cCal}$ = 8.5$^{+1.5}_{-1.4}\, \times\, $10$^{14}$~M$_{\odot}$. Initial observations using the SPT \citep{2011PASP..123..568C} revealed the cluster through its strong SZ effect signal, highlighting its significance as a unique object for studying galaxy formation and evolution, as well as for testing cosmological models \citep{2011ApJ...731...86F}. Further studies using the Atacama Large Millimeter/submillimeter Array (ALMA) and the Atacama Compact Array (ACA) provided evidence of a bimodal structure in the pressure distribution of the ICM, strongly suggesting the cluster is undergoing a major merger \citep{2021A&A...650A.153D}.
\newline
\newline
\cite{2021A&A...650A.153D} identified diffuse radio emission associated with the cluster at 943 MHz as part of the Evolutionary Map of the Universe \citep[EMU;][]{2011PASA...28..215N} Pilot Survey. \cite{2021A&A...650A.153D} found a spatial correlation between optical star-forming filaments seen in Hubble Space Telescope (HST) images and small-scale jet-like radio structure, suggesting a potential connection between the radio jet and the ongoing star formation activity. Additionally, weak lensing studies using HST data have provided precise measurements of the cluster's mass distribution, which is highly asymmetric, consisting of a primary clump and a smaller sub-clump located approximately 640 kpc to the west \citep{2019ApJ...887...76K,2021A&A...650A.153D}. 
\newline
\newline
In our analysis using the MeerKAT data, a low-resolution image revealed a radio halo spanning approximately $0.98\, \times\, 0.93$ Mpc in linear scale (see the top left panel of Figure \ref{fig:2a}). The diffuse emission, measured within $3\sigma$ contours, has an integrated flux density of $\mathrm{S_{1.28\,GHz} = 3.23 \pm 0.19\, mJy}$, corresponding to a k-corrected radio power at 1.4 GHz of $P_{1.4\,\mathrm{GHz}} = (2.64\, \pm\, 0.73) \times 10^{25}\, \mathrm{W\,Hz^{-1}}$.

\begin{table*}
\caption{Measured properties of all detected diffuse emissions within the sample.}
\label{table:3}
\begin{tabular}{ccccccc}
\hline
Cluster Name   & $R_{500c}$ & LAS                & LLS                & Classification & S$_{\nu}$            & $P_{1.4,\mathrm{GHz}}$         \\
(ACT-CL)       & (kpc)      & ($\prime$)         & (Mpc)              &                & (mJy)                & ($\times 10^{25}$ W Hz$^{-1}$) \\ \hline
J1137.8+0728   & 908.49     & $1.79 \times 1.38$ & $0.86 \times 0.66$ & RH             & $0.87\, \pm\, 0.08$  & $0.53\, \pm\, 0.14$            \\
J0546.6$-$5345 & 854.31     & $0.97 \times 0.90$ & $0.47 \times 0.44$ & Uncertain      & $0.42\, \pm\, 0.04$  & $0.30\, \pm\, 0.08$            \\
J2106.0$-$5844 & 939.07     & $1.99 \times 1.88$ & $0.98 \times 0.93$ & RH             & $3.23\, \pm\, 0.19$  & $2.64\, \pm\, 0.73$            \\
J1142.7+1527   & 829.06     & $1.96 \times 1.59$ & $0.97 \times 0.79$ & RH             & $1.47\, \pm\, 0.11$  & $1.38\, \pm\, 0.41$            \\
J0329.2$-$2330 & 951.05     & $2.17 \times 1.73$ & $1.08 \times 0.86$ & RH             & $3.45\, \pm\, 0.21$  & $3.55\, \pm\, 1.06$            \\
J0003.9+1642   & 786.96     & $1.96 \times 1.51$ & $0.99 \times 0.76$ & RH             & $1.43\, \pm\, 0.12 $ & $1.74\, \pm\, 0.55$            \\ \hline
\end{tabular}
\begin{minipage}{\textwidth}
\footnotesize
\textbf{Notes:} Columns: (1) ACT DR5 cluster name; (2) $R_{500c}$ of the mean mass over-density of the cluster 500 times the cosmic critical density at the cluster redshift; (3) largest angular size (LAS) of diffuse emission in arc-minutes; (4) largest physical extent of diffuse emission at cluster redshift in Mpc; (5) classification type: halo (RH) ; (6) total flux density in mJy; (7) $k$-corrected radio power at 1.4 GHz, assuming $\alpha = -1.3 \pm 0.4$.
\end{minipage}
\end{table*}

\subsubsection{ACT-CL J1142.7$+$1527}
ACT-CL J1142.7+1527 (hereafter, ACT-CL J1142, alternative name: 
MOO J1142+1527) is a massive galaxy cluster located at {\it z} = 1.19, with a mass of $M_{\rm 500cCal}$ = 6.3$^{+1.1}_{-1.0}\, \times\, $10$^{14}$ M$_{\odot}$ as reported in the ACT DR5 catalogue. Initially identified in the MaDCoWS survey \citep{2015ApJ...812L..40G,2019ApJS..240...33G}, its mass was later confirmed through SZ effect observations with the \textit{NIKA2}, which exhibited a decrement at a significance level of 13.2$\sigma$ \citep{2020ApJ...893...74R}. The \textit{NIKA2} and \textit{MUSTANG-2} SZ observations reveal the cluster to be an exceptionally massive, hot, and unrelaxed merging system, with \textit{MUSTANG-2} confirming this MaDCoWS cluster at a significance level of 20.9$\sigma$ \citep{2020ApJ...893...74R, 2020ApJ...902..144D}.
\newline
\newline
We observed diffuse radio emission at the centre of J1142 (see top right panel of Figure \ref{fig:2a}). The integrated flux density is $\mathrm{S\,_{1.28\,GHz} = 1.47 \pm 0.11\, mJy}$ from the low-resolution image, corresponding to a largest linear size (LLS) of $0.97\, \times\, 0.79$ Mpc, with the $k$-corrected radio power scaled to 1.4 GHz of $P_{1.4\,\mathrm{GHz}} = (1.38\, \pm\, 0.41) \times 10^{25}\, \mathrm{W\,Hz^{-1}}$. We classify the diffuse radio emission in J1142 as a radio halo since the emission is centrally located within the cluster, and displays a regular morphology that traces the X-ray emitting ICM. The brightest radio source ($\sim200\, \rm{kpc}$ away from the SZ peak) flux density, which is $\mathrm{S\,_{1.28\,GHz} = 37.27 \pm\, 1.85\, mJy}$, is consistent with the one found in FIRST \citep[37 mJy,][]{2015ApJ...801...26H}. \citet{2020ApJ...893...74R} suggested that J1142 might be undergoing a merger. Based on a multi-wavelength analysis including \textit{Chandra} X-ray and \textit{NIKA2} SZ data, they further show that the central region of the cluster is hosting relatively cool gas, reminiscent of what could be the remnant cool core from one of the interacting systems.

\begin{figure*}
\centering
\begin{tabular}{cc}
    \includegraphics[width=0.48\textwidth]{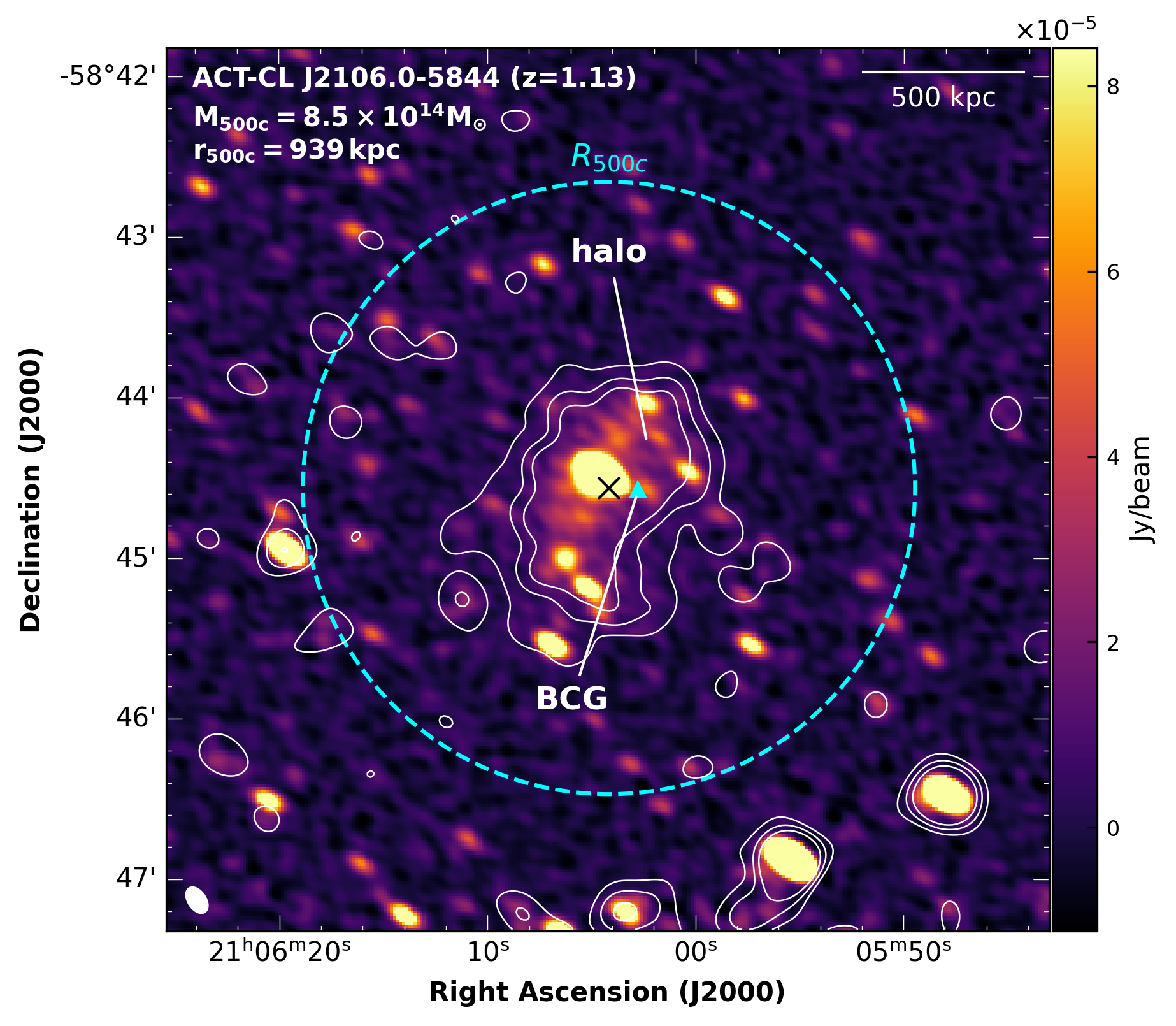} &
    \includegraphics[width=0.48\textwidth]{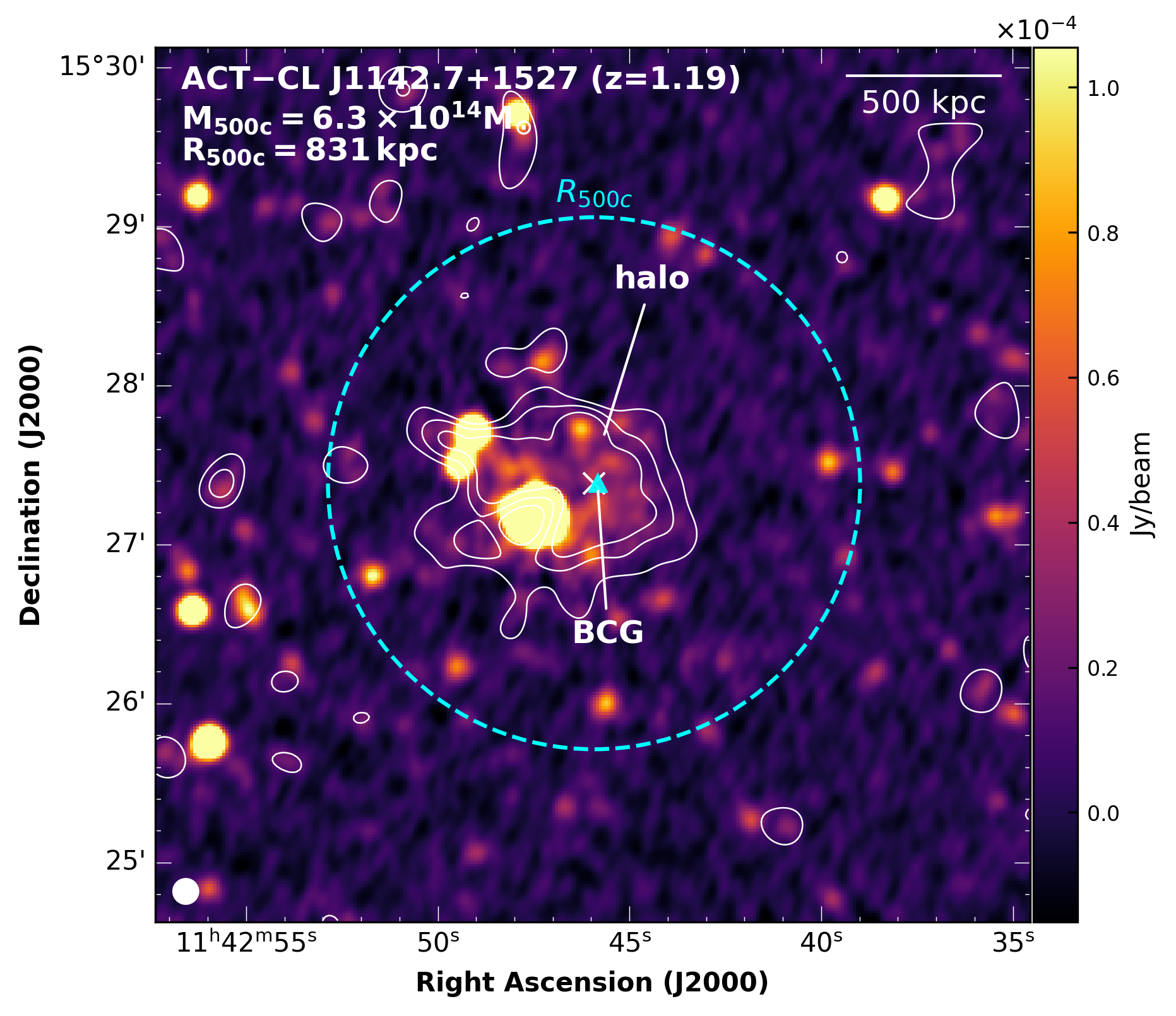} \\
    \includegraphics[width=0.48\textwidth]{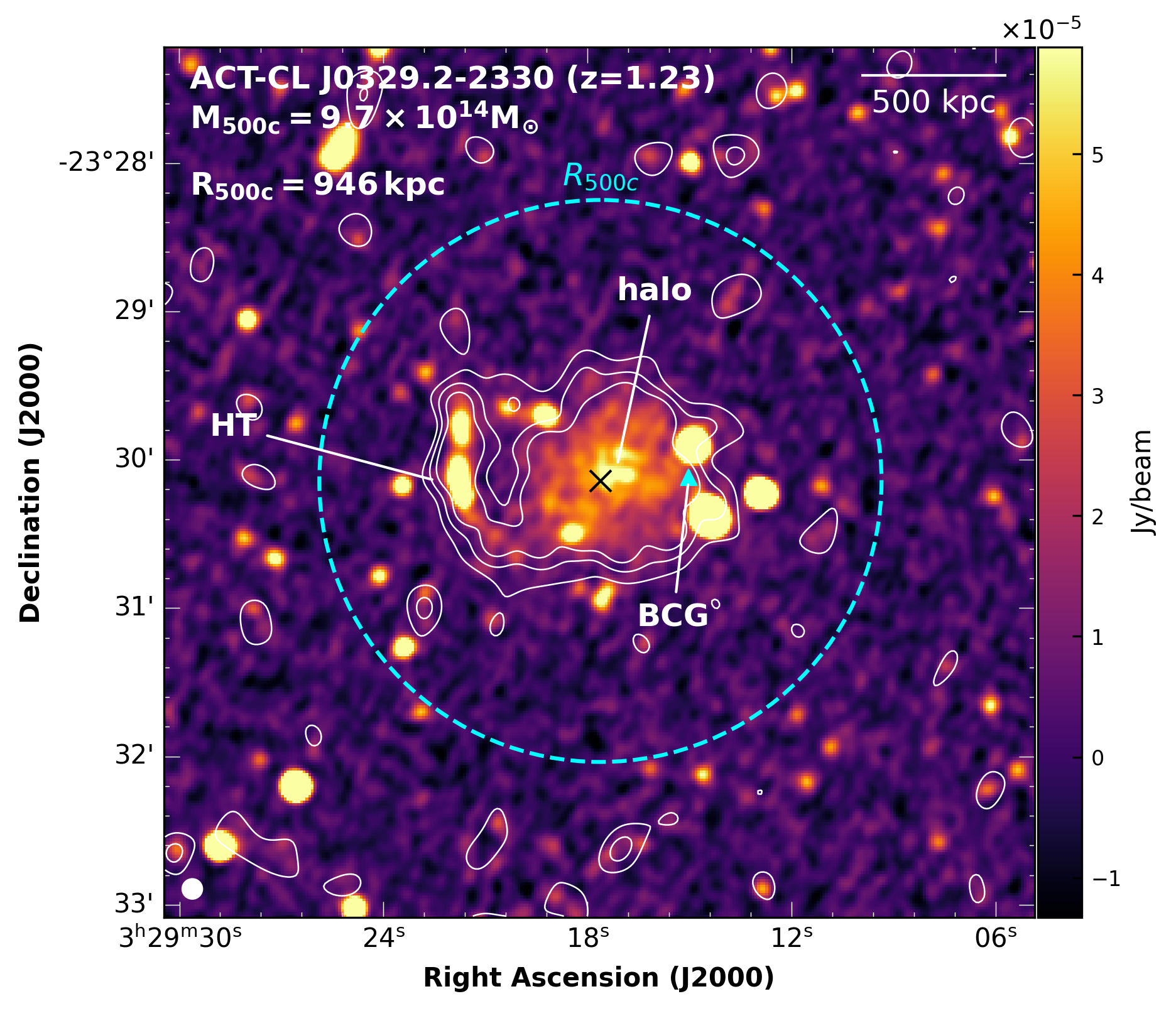} &
    \includegraphics[width=0.48\textwidth]{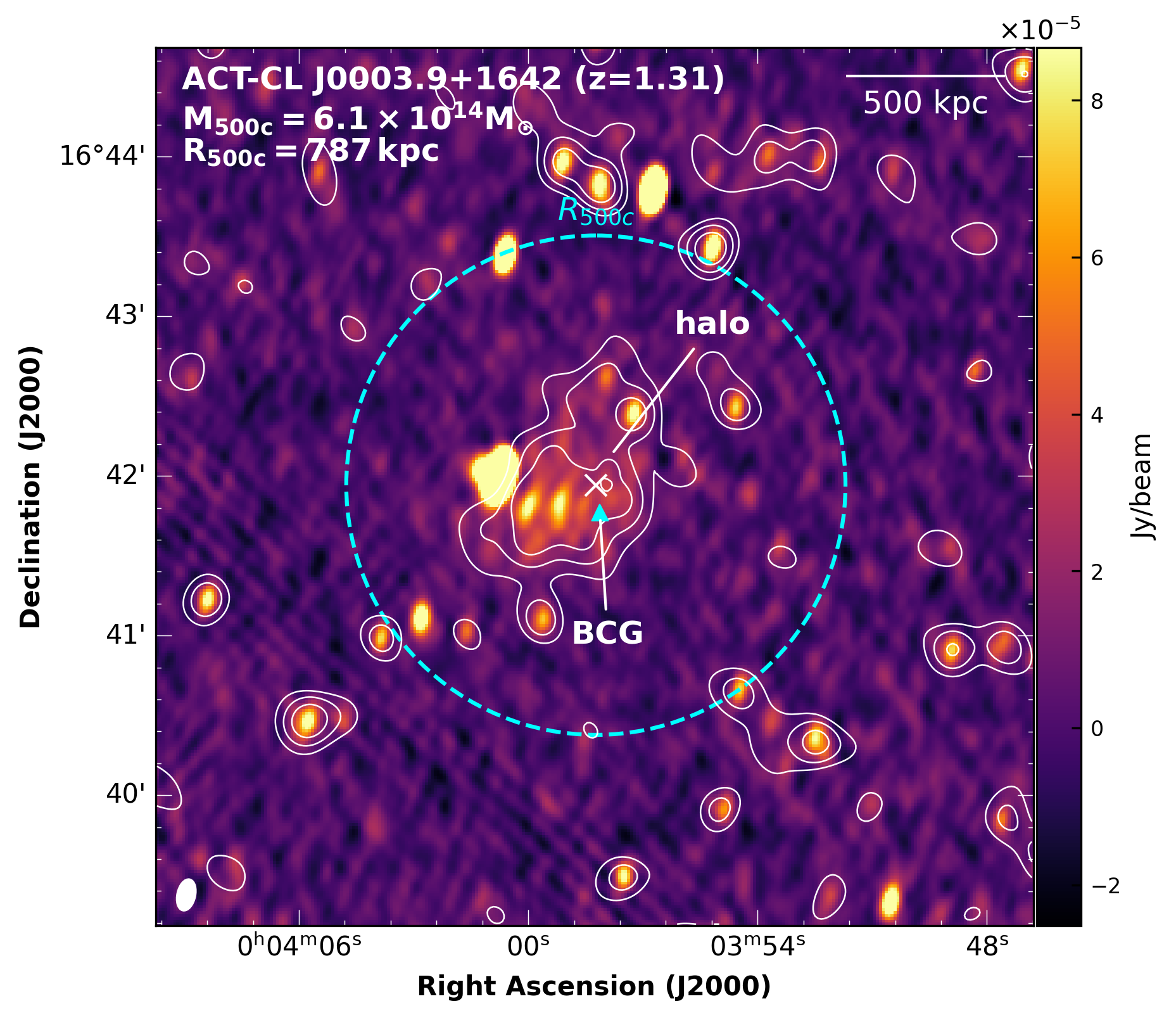}
\end{tabular}
\caption{Full-resolution MeerKAT $L$-band images of clusters hosting radio halos. The superimposed contours are from the low-resolution source-subtracted images, with contour levels set at $\mathrm{\sigma_{LR} \times [3, 6, 10]}$. The projected physical scale at each cluster's redshift is indicated in the top right of each panel. The synthesized beam, shown as a filled white ellipse or circle, is displayed in the lower left corner of each panel, and the location of the ACT SZ peak is marked by the cross (black/white). The cyan triangle indicates the position of the BCG, adopted from the ACT DR5 catalogue \citep{2021ApJS..253....3H}. \textit{Top left:} ACT-CL J2106.0-5844 cluster, the full-resolution image has a local rms of $1\sigma\, = 4.5\, \mathrm{\mu Jy/beam}$, while the low-resolution image used for the contours has a local rms of $1\sigma\, = 6.5\, \mathrm{\mu Jy/beam}$. \textit{Top right:} ACT-CL J1142.7+1527 cluster, the full-resolution image has a local rms of $1\sigma\, = 5.4\, \mathrm{\mu Jy/beam}$, while the low-resolution image used for the contours has a local rms of $1\sigma\, = 7.2\, \mathrm{\mu Jy/beam}$. \textit{Bottom left:} ACT-CL J0329.2-2330 cluster, the full-resolution image has a local rms of $1\sigma\, = 5.0\, \mathrm{\mu Jy/beam}$, while the low-resolution image used for the contours has a local rms of $1\sigma\, = 6.1\, \mathrm{\mu Jy/beam}$. \textit{Bottom right:} ACT-CL J0003.9+1642 cluster, the full-resolution image has a local rms of $1\sigma\, = 7.4\, \mathrm{\mu Jy/beam}$, while the low-resolution image used for the contours has a local rms of $1\sigma\, = 9.2\, \mathrm{\mu Jy/beam}$.}
\label{fig:2a}
\end{figure*}

\begin{figure*}
\ContinuedFloat
\centering
\begin{multicols}{2}
    \includegraphics[width=0.48\textwidth]{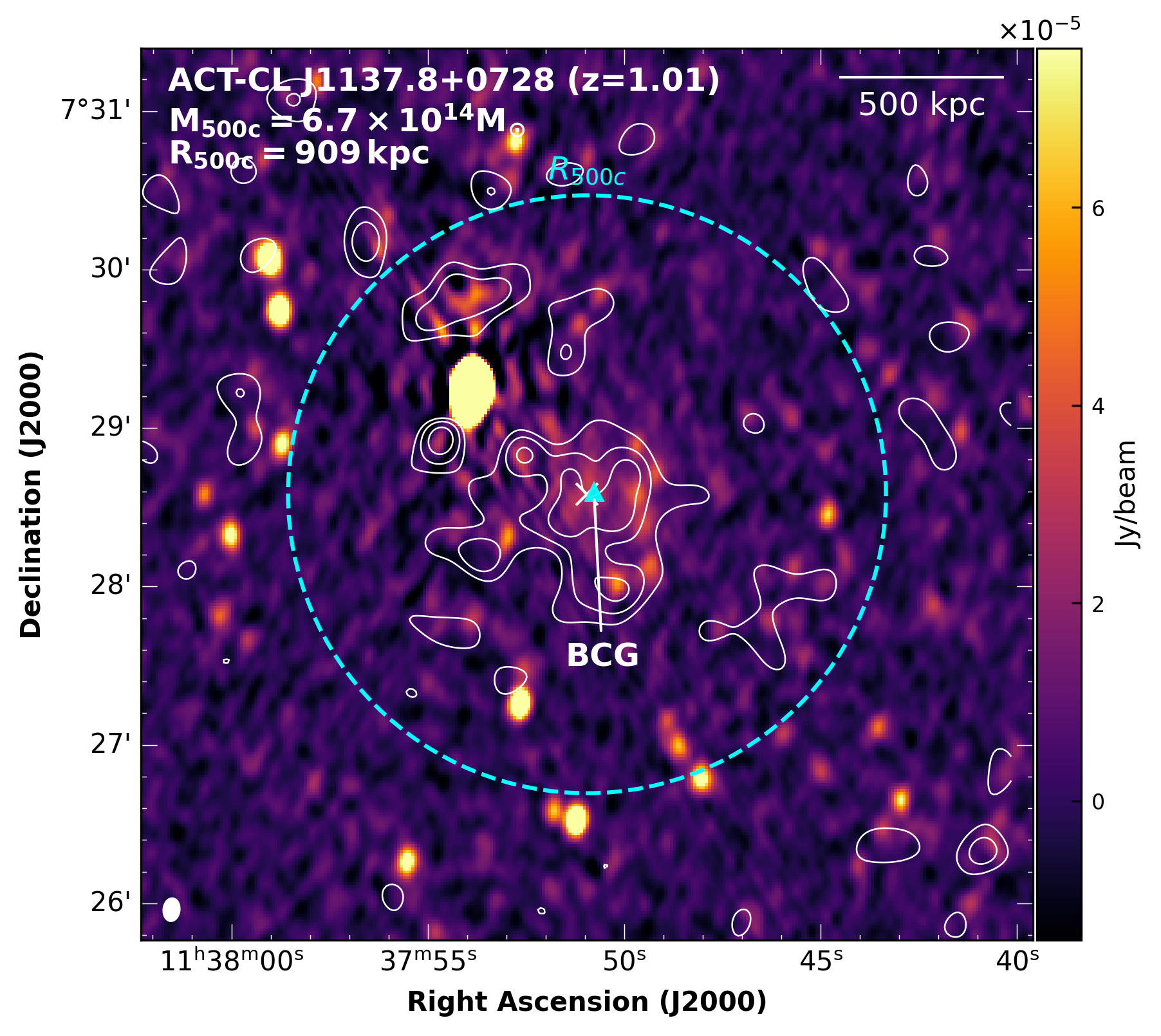}\par 
    \includegraphics[width=0.48\textwidth]{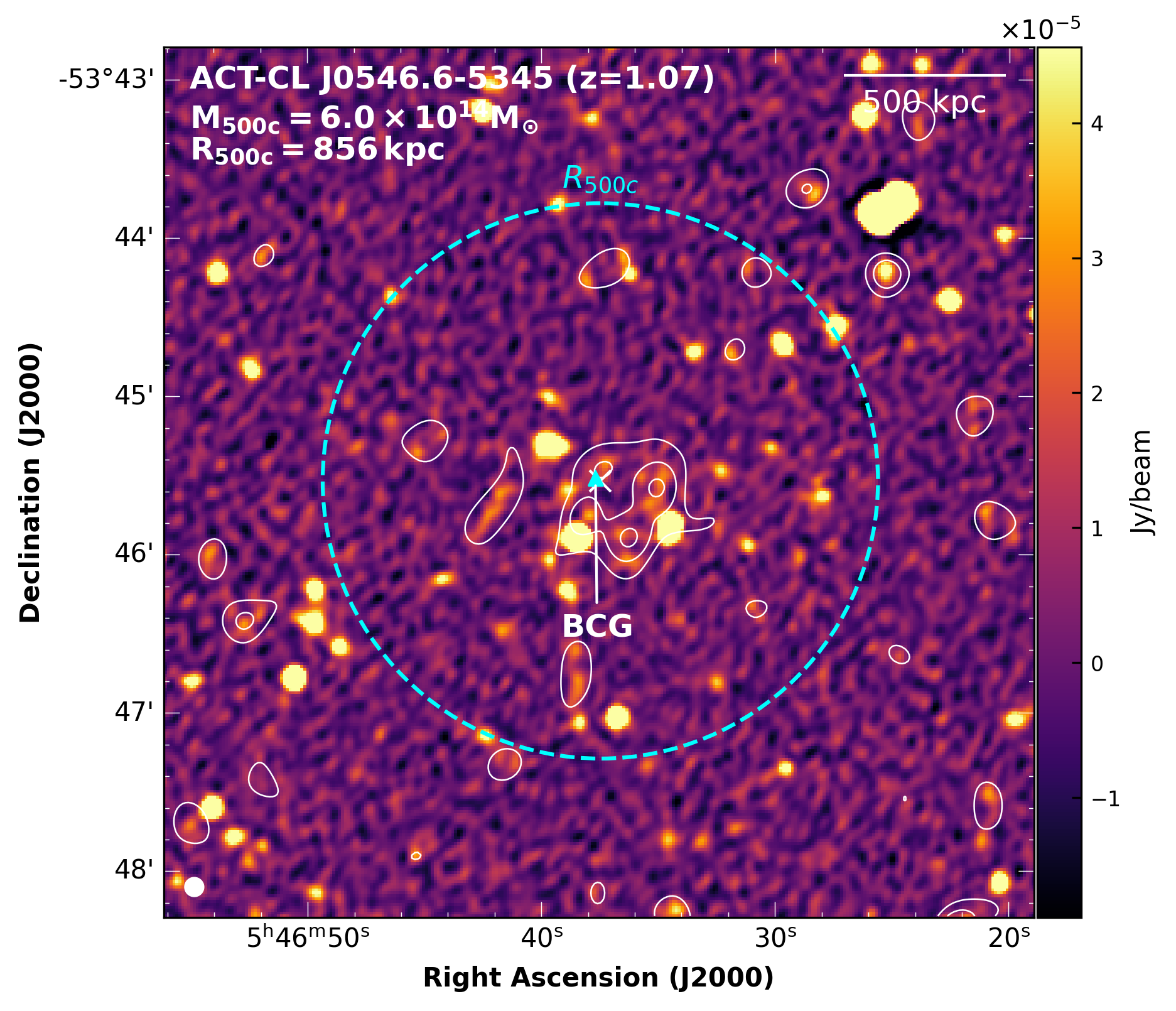}\par 
\end{multicols}
\caption{Full-resolution MeerKAT $L$-band images of clusters hosting radio halos (continued). The superimposed contours are from the low-resolution source-subtracted images, with contour levels set at $\mathrm{\sigma_{LR} \times [3, 6, 10]}$. The projected physical scale at each cluster's redshift is indicated in the top right of each panel. The synthesized beam, shown as a filled white ellipse or circle, is displayed in the lower left corner of each panel, and the location of the ACT SZ peak is marked by the white cross. The cyan triangle indicates the position of the BCG, adopted from the ACT DR5 catalogue \citep{2021ApJS..253....3H}. \textit{Left:} ACT-CL J1137.8+0728 cluster, the full-resolution image has a local rms of $1\sigma\, = 5.4\, \mathrm{\mu Jy/beam}$, while the low-resolution image used for the contours has a local rms of $1\sigma\, = 7.5\, \mathrm{\mu Jy/beam}$. \textit{Right:} ACT-CL J0546.6-5345 cluster, the full-resolution image has a local rms of $1\sigma\, = 6.1\, \mathrm{\mu Jy/beam}$, while the low-resolution image used for the contours has a local rms of $1\sigma\, = 9.4\, \mathrm{\mu Jy/beam}$.}
\label{fig:2b}
\end{figure*}

\subsubsection{ACT-CL J0329.2-2330}
ACT-CL J0329.2-2330, hereafter ACT-CL J0329, with a mass of $M_{\rm 500cCal}$ =  9.7$^{+1.7}_{-1.6}\, \times\, $10$^{14}$ M$_{\odot}$ and {\it z} = 1.23, is among the highest-redshift galaxy clusters confirmed to host a radio halo \citep{2025A&A...698L..17S}. J0329 was detected by both the ACT and SPT surveys, both of which use the SZ effect to identify distant, massive clusters \citep{2021ApJS..253....3H,2020ApJS..247...25B}.
\newline
\newline
Our analysis reveals a prominent radio halo located in the central region of the cluster. To the east, we observe a nearby elongated source that connects to the halo emission (see the left bottom panel of Figure \ref{fig:2a}). This source exhibits a morphology consistent with a head-tail (HT) radio galaxy and its flux density is $0.65\, \pm 0.01\, \rm{mJy}$. The measured flux density of the halo is $\mathrm{S\,_{1.28\,GHz} = 3.45 \pm 0.21\, mJy}$, excluding the tailed emission from the elongated source. The angular size of the halo is $2.17' \times 1.73'$, corresponding to a linear size of $1.08\, \times\, 0.86$ Mpc. The $k$-corrected radio power at 1.4 GHz is $P_{1.4\,\mathrm{GHz}} = (3.55\, \pm\, 1.06)\, \times\, 10^{25}\, \mathrm{W\,Hz^{-1}}$. 
\newline
\newline
The findings are consistent with those of \citet{2025A&A...698L..17S}, who also detected a radio halo in ACT-CL J0329.2-2330 with comparable flux density and angular/linear size characteristics.  While the same MeerKAT $L$-band data were used in both studies, this work represents an independent analysis conducted using the \texttt{Oxkat} pipeline, distinct from the \texttt{CARACal}\footnote[12]{\url{https://github.com/caracal-pipeline/caracal}} pipeline \citep{2020ascl.soft06014J} employed by \citet{2025A&A...698L..17S}. In addition, the elongated source to the east of the cluster’s centre is identified as a HT radio galaxy. This HT galaxy provides valuable insights into the cluster’s dynamical state and the possible merger history. The presence of such sources is often associated with dense ICM conditions and shocks generated by ongoing dynamical activity, further supporting the idea that the cluster has experienced significant merger-driven interactions \citep{2025A&A...698L..17S}.

\subsubsection{ACT-CL J0003.9$+$1642}

Located at a redshift {\it z} = 1.31, ACT-CL J0003.9+1642 (hereafter ACT-CL J0003) is a galaxy cluster with a mass of $M_{\rm 500cCal}$ = 6.1$^{+1.1}_{-1.0} \times $10$^{14}\, \mathrm{M_{\odot}}$, initially identified through the ACT’s SZ effect measurements \citep{2021ApJS..253....3H}. Our analysis of MeerKAT data reveals diffuse radio emission in the form of a radio halo concentrated near the cluster’s centre (see bottom right panel of Figure \ref{fig:2a}), spanning an angular size of $1.96' \times 1.51'$, which corresponds to a linear size of $0.99 \times 0.76$ Mpc. The integrated flux density of the diffuse emission is measured to be $\mathrm{S\,_{1.28\,GHz} = 1.43 \pm 0.12\, mJy}$, translating to a $k$-corrected radio power of $P_{1.4\,\mathrm{GHz}} = (1.74\, \pm\, 0.55) \times 10^{25}\, \mathrm{W\,Hz^{-1}}$. 
\newline
\newline
The brightest radio source in the image, located $\sim\! 38^{\prime \prime}$ east of the halo's central peak (projected distance  $\sim 317\, \rm{kpc}$), shows a flux density of $2.83\, \pm\, 0.14$ mJy. This offset compact source is distinct from the BCG visible in optical counterparts (Figure \ref{fig:8}). It is spatially coincident with a point-like, pink-coloured optical source, suggesting it is likely a foreground star unrelated to the diffuse emission. The BCG itself is located $<20^{\prime \prime}$ from SZ peak position in the south. The presence of the central radio halo nevertheless indicates merger-driven turbulence \citep{2012A&ARv..20...54F,2019SSRv..215...16V}, potentially from a recent minor merger or line-of-sight accretion event that preserved the overall cluster symmetry.
\subsection{Faint Diffuse Emission}
\label{sec:Faint Diffuse Emission Detected}

\subsubsection{ACT-CL J1137.8+0728}

The galaxy cluster ACT-CL J1137.8+0728 (hereafter, ACT-CL J1137), located at a redshift of {\it z} = 1.01, has a total estimated mass of $M_{\rm 500cCal}$ = 6.7$^{+1.5}_{-1.4} \times $10$^{14}$ M$_{\odot}$. The initial full-resolution MeerKAT image showed no evidence of diffuse emission, limited by the low surface brightness of the ICM. However, after applying point-source subtraction and re-imaging with lower resolution techniques, faint diffuse radio emission emerged, reaching a significance threshold of $3\sigma$.
\newline
\newline
The measured flux density of diffuse emission is $\mathrm{S\,_{1.28\,GHz} = 0.87 \pm 0.08\, mJy}$, and the $k$-corrected radio power of the halo is $P_{1.4\,\mathrm{GHz}} = (0.53\, \pm\, 0.14) \times 10^{25}\, \mathrm{W\,Hz^{-1}}$. The diffuse emission extends across a region that is centred on the core of the cluster, as shown in Figure \ref{fig:2b}. The cyan dashed circle represents $R_{500c}\, =\, 908$ kpc, and the detected emission lies within this region. The brightest radio source is located approximately $60^{\prime \prime}$ north-east of the SZ peak centre, with a flux density of $\mathrm{S\,_{1.28\,GHz} = 39.23 \pm 1.96\, mJy}$. Despite the use of direction-dependent calibration, residual artefacts from this bright source are visible in the image and extend toward the central region of the cluster. While these artefacts are largely suppressed, they may contribute marginally to the measured flux of the diffuse emission.

\subsubsection{ACT-CL J0546.6-5345}

ACT-CL J0546.6-5345 (hereafter, ACT-CL J0546, also known as SPT$-$CL J0546.6-5345) is a galaxy cluster located at {\it z} = 1.07, with an estimated mass of $M_{\rm 500cCal}$ = 6.0$^{+1.1}_{-1.0} \times $10$^{14}$ M$_{\odot}$. It was initially detected through its SZ effect signature in a survey conducted using the SPT. The initial detection was reported by \cite{2009ApJ...701...32S}, and subsequent confirmation and analysis were conducted by \cite{2010ApJ...721...90B}.
\newline
\newline
Using MeerKAT observations, faint diffuse radio emission was identified within the cluster. Although the emission was not visible in the full-resolution image, the subtraction of compact sources in the low-resolution data revealed the diffuse component at the $3\sigma$ level (see Figure \ref{fig:2b}). The measured flux density of the emission is $\mathrm{S\,_{1.28\,GHz} = 0.42 \pm 0.04\, mJy}$, which yields a $k$-corrected radio power of $P_{1.4\,\mathrm{GHz}} = (0.30\, \pm\, 0.08) \times 10^{25}\, \mathrm{W\,Hz^{-1}}$. ACT-CL J0546.6–5345 exhibits faint, patchy radio emission tracing individual galaxies rather than the ICM. While statistically significant at 3$\sigma$, the emission lacks the spatial extent ($\ge$0.5 Mpc) and central concentration required for a radio halo classification \citep{2010ApJ...721L..82C}. This emission likely originates from AGN activity or fossil plasma from past minor mergers, consistent with its near-relaxed X-ray morphology (see Figure \ref{fig:6}). Given the absence of a clear, centrally located, and Mpc-scale emission feature, we classify the radio emission in ACT-CL J0546 as uncertain.

\subsection{Dynamical State of the Galaxy Clusters}
Galaxy clusters are dynamically evolving systems, continuously accreting matter and undergoing mergers throughout cosmic time. While they are often categorised as relaxed or disturbed for observational convenience, these labels represent endpoints on a spectrum of dynamical states rather than discrete classes. Relaxed clusters typically have a symmetrical distribution of galaxies and intracluster gas around their BCG, indicating they have reached a state of approximate virial equilibrium. In contrast, dynamically disturbed clusters deviate significantly from virial equilibrium, often displaying substructure, multiple bright galaxies, and irregular X-ray morphologies. The presence and morphology of diffuse radio emission are closely linked to the dynamic state, as mergers and disturbances can inject turbulence into the ICM, leading to the (re)acceleration of cosmic ray electrons \citep{2010ApJ...721L..82C}.
\newline
\newline
The \textit{Chandra} X-ray data were obtained from the \textit{Chandra} Data Archive\footnote[13]{\url{http://cda.harvard.edu/chaser/}} (ObsIDs listed in Figures \ref{fig:3}-\ref{fig:5}). Observations were conducted with ACIS-I in VFAINT mode, spanning 0.7–7.0 keV. Data reduction followed standard procedures using \textit{Chandra} Interactive Analysis of Observations \citep[\texttt{CIAO} v4.15,][]{2006SPIE.6270E..1VF}\footnote[14]{\url{https://cxc.cfa.harvard.edu/ciao/download/}} and calibration database \citep[\texttt{CalDB} v4.12.0,][]{2007ChNew..14...33G}.\footnote[15]{\url{https://cxc.cfa.harvard.edu/caldb/}} Of the six clusters in our sample, five have archival \textit{Chandra} X-ray data, and ACT-CL J1137.8+0728 lacks X-ray observations. To assess the dynamical state of each cluster, we adopt the morphological classification scheme defined by \citet{2020MNRAS.497.5485Y} and \citet{2022MNRAS.513.3013Y}, which uses a combination of X-ray morphological indicators.

\subsubsection*{ACT-CL J2106.0-5844}

\begin{figure*}
\begin{multicols}{2}
    \includegraphics[width=0.48\textwidth]{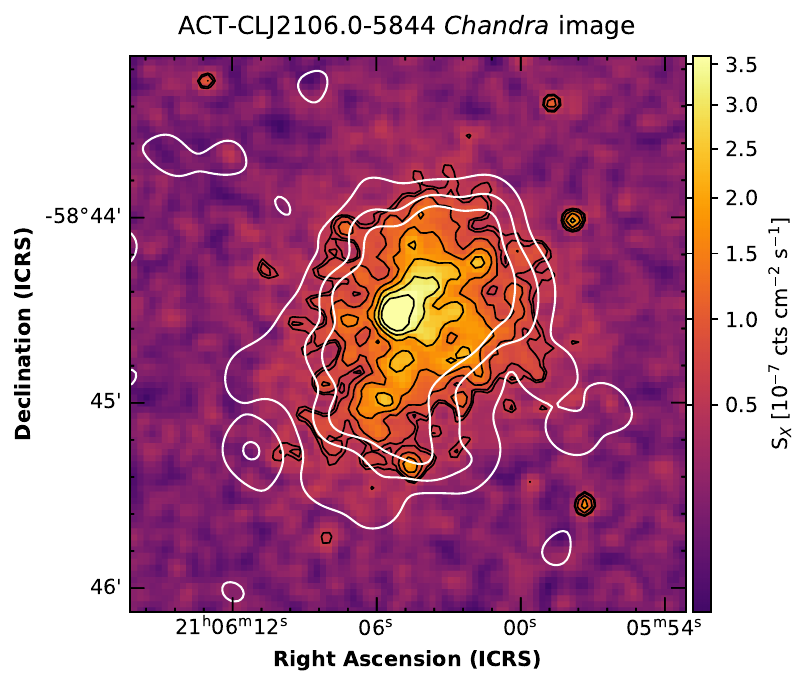}\par 
    \includegraphics[width=0.48\textwidth]{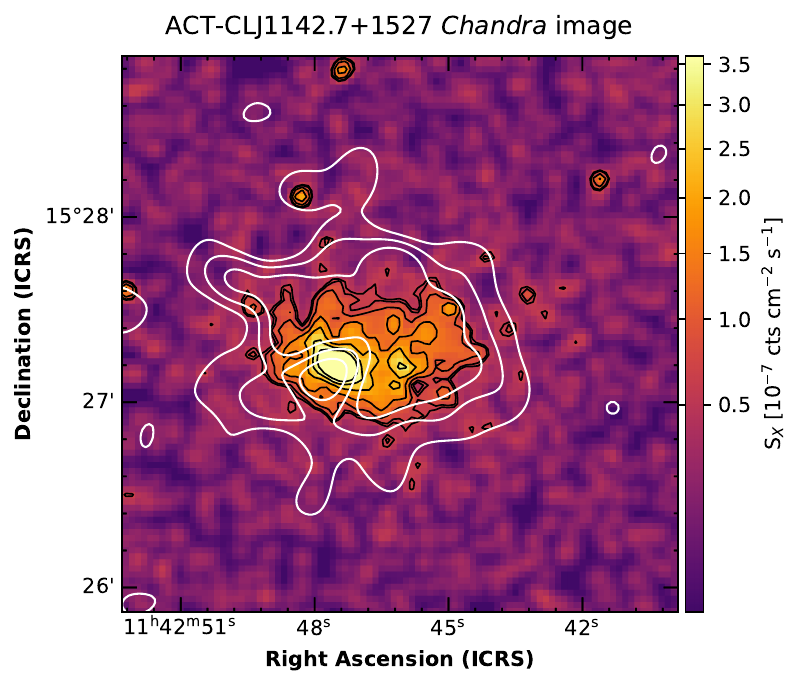}\par 
    \end{multicols}

\caption{\textit{Chandra} exposure-corrected images were created using the $0.7-7.0$ keV energy band and observed using ACIS-I in VFAINT mode exposure-corrected images. For the purpose of display, the images were $4\times4$ pixel binned to 1.968$^{\prime\prime}$ resolution and then smoothed with a Gaussian kernel of 1 pixel full-width half-maximum (FWHM). The black contours trace the X-ray surface brightness at levels of $[0.6, 0.69, 0.98, 1.5, 2.1, 3.0, 4.0]\, \times\, 10^{-7} \rm{cts\, cm^{-2}\, s^{-1}}$. The white superimposed contours from the MeerKAT LR image are shown at levels of $\mathrm{\sigma \times [3, 6, 10]}$, where $1\sigma$ corresponds to $6.5\, \mathrm{\mu Jy/beam}$ of the J2106 cluster (left) and $7.2\, \mathrm{\mu Jy/beam}$ for J1142 cluster (right). The left panel shows the J2106 cluster \textit{Chandra} image, generated from observation IDs 12180 and 12189 (PI: S. Murray and G. Garmire) with an exposure time of 24.72 and 48.13 ksec, respectively. The right panel shows the J1142 cluster \textit{Chandra} image, generated from observation ID 18277 (PI: S. Stanford) with an exposure time of 46.95 ksec.}
\label{fig:3}
\end{figure*}

The \textit{Chandra} X-ray image of ACT-CL J2106.0-5844 (Figure \ref{fig:3}, left panel) shows clear signs of a dynamically disturbed morphology, with asymmetric emission and a tail-like feature extending to the northwest, which are indicative of an ongoing merger. Although the BCG lies near the X-ray peak (offset by $<5$ kpc), this spatial alignment does not preclude merger activity. The combination of the disturbed X-ray structure and the presence of a central radio halo suggests active turbulence in the ICM, which is commonly associated with merging systems. The radio halo broadly traces the thermal X-ray emission and points to merger-driven processes that re-accelerate relativistic particles \citep{2014IJMPD..2330007B,2019SSRv..215...16V}. Constraining the geometry of the merger more precisely would require additional probes, including galaxy velocity distributions, detailed X-ray temperature maps, and the spatial alignment of substructures.

\subsubsection*{ACT-CL J1142.7+1527}
The \textit{Chandra} X-ray image reveals a moderately elongated morphology with asymmetric substructure in the core region (see the right panel of Figure \ref{fig:3}). The BCG is spatially coincident with the SZ peak but offset by $\sim200\, \rm{kpc}$ from both the X-ray emission peak and the large-scale ICM centroid. This triple offset, BCG/SZ peak, X-ray emission peak, and ICM centroid, indicates ongoing merger activity, as relaxed clusters typically exhibit alignment between these components. The MeerKAT radio halo shows (Figure \ref{fig:2a}, right panel) a centroid offset of $10-20\, \rm{arcsec}$ ($\sim150-300\, \rm{kpc}$) from the BCG, aligning more closely with the ICM centroid. This spatial relationship implies the halo traces turbulent motions in the bulk ICM rather than being directly tied to the BCG’s position.

\subsubsection*{ACT-CL J0329.2-2330}
The \textit{Chandra} X-ray image reveals a moderately elongated morphology along the NE-SW direction with asymmetric substructure in the core region (see the left panel of Figure \ref{fig:4}). This morphology suggests a dynamically disturbed state. The X-ray image exhibits significant asymmetry and an irregular shape. Furthermore, the radio contours indicate that the radio halo emission is generally co-located with the X-ray emission, providing evidence for recent or ongoing merger activity that likely generates the observed turbulence and diffuse radio emission.

\subsubsection*{ACT-CL J0003.9+1642}
\begin{figure*}
\begin{multicols}{2}
\includegraphics[width=0.48\textwidth]{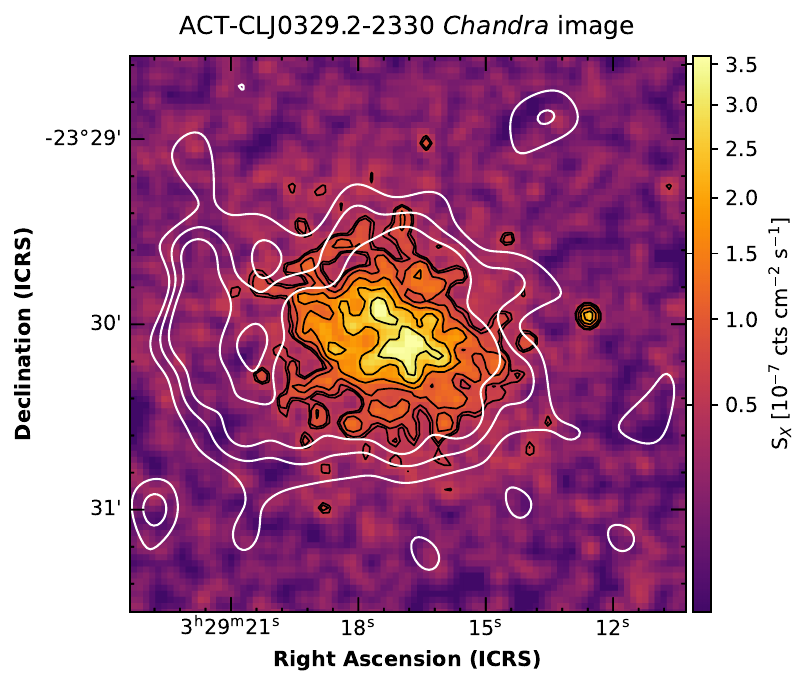}\par 
\includegraphics[width=0.48\textwidth]{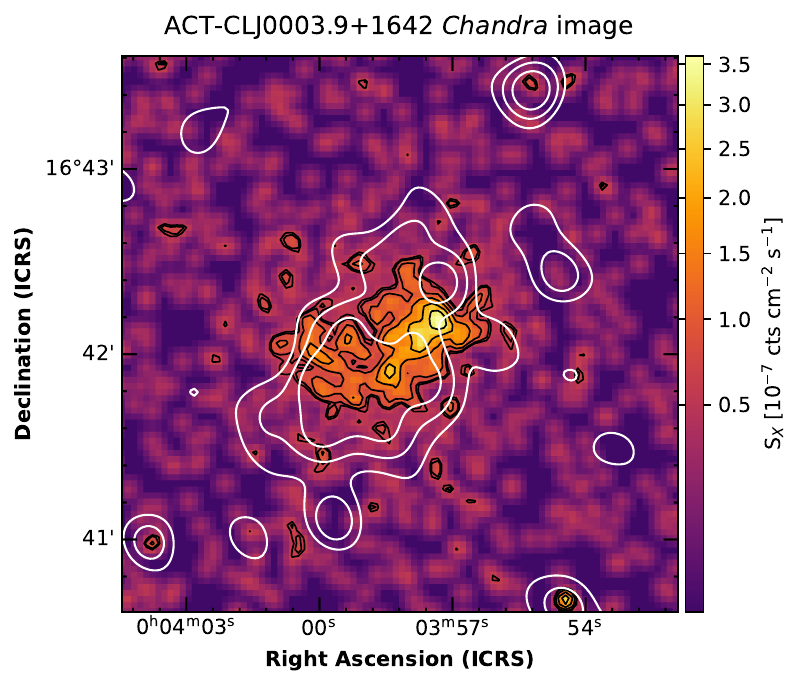}\par 
\end{multicols}
\caption{\textit{Chandra} exposure-corrected images were created using the $0.7-7.0$ keV energy band and observed using ACIS-I in VFAINT mode exposure-corrected images. For the purpose of display, the images were $4\times4$ pixel binned to 1.968$^{\prime\prime}$ resolution and then smoothed with a Gaussian kernel of 1 pixel full-width half-maximum (FWHM). The black contours trace the X-ray surface brightness at levels of $[0.6, 0.69, 0.98, 1.5, 2.1, 3.0, 4.0]\, \times\, 10^{-7} \rm{cts\, cm^{-2}\, s^{-1}}$. The white superimposed contours from the MeerKAT LR image are shown at levels of $\mathrm{\sigma \times [3, 6, 10]}$, where $1\sigma$ corresponds to $6.1\, \mathrm{\mu Jy/beam}$ of the J0329 cluster (left) and $9.2\, \mathrm{\mu Jy/beam}$ for J0003 cluster (right). The left panel shows the J0329 cluster \textit{Chandra} image, generated from observation IDs 18282 and 18882 (PI: L. Bleem) with an exposure time of 26.14 and 21.37 ksec, respectively. The right panel shows the J0003 cluster \textit{Chandra} image, generated from observation ID 26935 (PI: A. Flores) with an exposure time of 19.80  ksec.}
\label{fig:4}
\end{figure*}

The X-ray emission from ACT-CL J0003.9+1642 (see the right panel of Figure \ref{fig:4}) displays a notably irregular and elongated morphology, with some possible substructure evident. The BCG is located within 20 arcseconds of the SZ peak toward the south. Despite this proximity, the distorted X-ray morphology and the presence of a central radio halo suggest that the cluster is dynamically disturbed. These features indicate that merger-driven turbulence is likely impacting the distribution of the ICM, possibly as a result of ongoing or recent merger activity.

\subsubsection*{ACT-CL J0546.6-5345}
\begin{figure}

\includegraphics[width=0.48\textwidth]{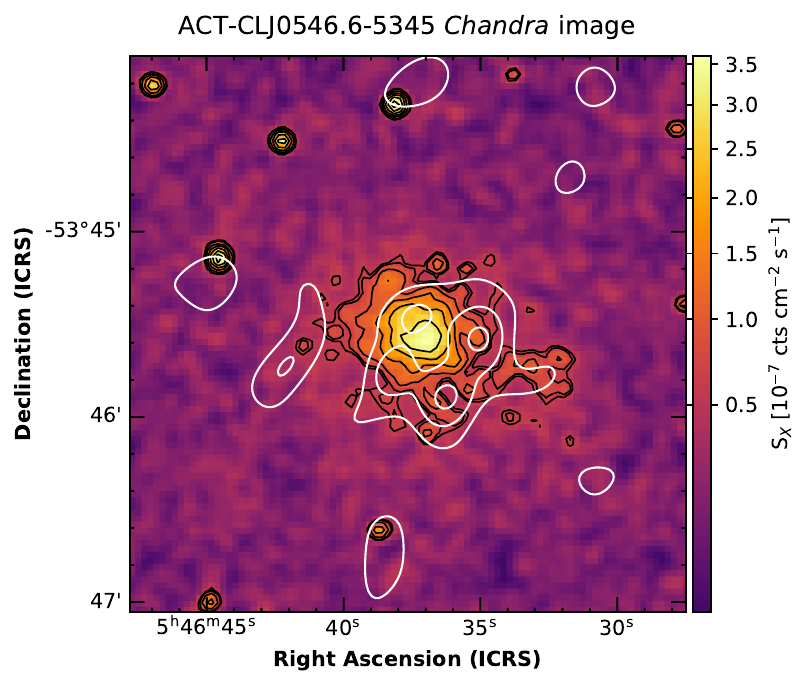}\par 
\caption{\textit{Chandra} exposure-corrected image is created using the $0.7-7.0\, \rm{keV}$ energy band and observed using ACIS-I in VFAINT mode exposure-corrected images. For the purpose of display, the image was $4\times4$ pixel binned to 1.968$^{\prime\prime}$ resolution and then smoothed with a Gaussian kernel of 1 pixel FWHM. The J0546 cluster \textit{Chandra} image, generated from observation IDs 9332, 10851, 10864, 9336 and 11739 (PI: G. Garmire and J. Mohr) with an exposure time of 14.88, 8.52, 5.76, 28.17, and 12.78 ksec, respectively. The black contours trace the X-ray surface brightness at levels of $[0.6, 0.69, 0.98, 1.5, 2.1, 3.0, 4.0]\, \times\, 10^{-7} \rm{cts\, cm^{-2}\, s^{-1}}$, and the white superimposed contours from the MeerKAT LR image have contour levels of  $\mathrm{\sigma \times [3, 6, 10]}$, with $1\sigma = 9.4\, \mathrm{\mu Jy/beam}$.}
\label{fig:5}
\end{figure}

ACT-CL J0546.6-5345 presents an intriguing case. The \textit{Chandra} X-ray image (see Figure \ref{fig:5}) shows a nearly relaxed morphology with a central peak of X-ray emission, suggesting the ICM is quite symmetrical and possibly nearly virialised.  However, we note a clear spatial offset between the peak of the X-ray emission and the centroid of the radio emission. Such an offset is atypical for relaxed clusters and may signal ongoing dynamical activity or projection effects. This cluster was classified as dynamically disturbed by \citet{2020MNRAS.497.5485Y} and \citet{2022MNRAS.513.3013Y}, based on morphology index. Morphology index, $\delta$, is based on X-ray surface brightness fluctuations where $\delta \ge 0$ indicates disturbed systems and $\delta < 0$ indicates relaxed clusters \citep{2020MNRAS.497.5485Y}. For ACT-CL J0546.6-5345, the morphology index $\delta$ is above zero, implying disturbance. Additionally, velocity dispersion analysis of cluster galaxies from \citet{2016MNRAS.461..248S} reveals significant dynamical disturbance via the Dressler-Shectman (DS) test \citep{1988AJ.....95..985D}, inconsistent with a relaxed system. This discrepancy between X-ray and optical data could indicate a line-of-sight merger, where the collision axis aligns with our viewing direction \citep{2012MNRAS.420.2120M}. In such geometry, the X-ray morphology remains symmetric due to projection effects, while galaxy kinematics retain merger-induced velocity anisotropies. 
\newline
\newline
The radio emission from MeerKAT data appears to be tracing individual radio galaxies within the cluster rather than a diffuse radio halo. The lack of a clear radio halo is consistent with one possible interpretation where an edge-on (line-of-sight) merger geometry reduces the observable turbulent re-acceleration, limiting radio halo formation or its detectability \citep{2012MNRAS.420.2120M}. Nonetheless, it is also possible that this cluster simply does not host a radio halo, or that any halo is too faint to be detected at the current MeerKAT $L$-band sensitivity \citep[see][for occurrence rates and detection thresholds at higher redshifts]{2013ApJ...777..141C,2021A&A...647A..51C}. The system likely represents an early-stage merger where bulk motions dominate over shock-driven turbulence, rather than a post-merger relaxing system.

\section{Discussion}
\label{sec:Discussion}
Our analysis of six most massive high-redshift clusters selected based on their SZ mass reveals a 83$\% $ detection rate of diffuse radio emission, providing further insights into the dynamical states and ICM processes of these clusters. By plotting the scaling relation $P\mathrm{_{1.4\,GHz}}$ versus $M\mathrm{_{500}^{unc}}$ (see Figure \ref{fig:6}), we find that the measured radio halo powers align broadly with the established trends observed in clusters with lower redshifts. We use uncorrected SZ masses ($M\mathrm{_{500}^{unc}}$) for consistency with the $P\mathrm{_{1.4\,GHz}}$ $-$ $M\mathrm{_{500}}$ correlation literature \citep{2013ApJ...777..141C,2023A&A...680A..30C}. These masses derive directly from the Universal Pressure Profile scaling relation \citep{2010A&A...517A..92A} without weak lensing (WL) calibration adjustments, avoiding systematic uncertainties from mass bias corrections \citep{2021ApJS..253....3H}.

\subsection{Characteristics of Prominent Radio Halos}
\label{sec:Characteristics of Prominent Radio Halos}
The prominent radio halos detected in this study exhibit key characteristics that align with expectations for diffuse emission in merging galaxy clusters. These halos are extended, low-surface-brightness features that trace the thermal ICM. Their spatial extent ranges from approximately 0.97 Mpc to 1.08 Mpc, consistent with typical sizes observed in lower-redshift clusters. The measured flux densities of the halos span 1.43 mJy to 3.45 mJy at 1.28 GHz, corresponding to $k$-corrected radio powers scaled to 1.4 GHz in the range $1.38 \times 10^{25}\, \mathrm{W\,Hz^{-1}}$ to $3.55 \times 10^{25}\, \mathrm{W\,Hz^{-1}}$.
\newline
\newline
The radio halos are spatially coincident with X-ray emission from the ICM, as confirmed by \textit{Chandra} observations. This co-location aligns with the disturbed X-ray morphologies, characterised by asymmetries and substructures observed in these clusters, which indicate dynamically active systems. Together, these features support the model where merger-driven turbulence re-accelerates cosmic ray electrons and amplifies magnetic fields, generating the observed synchrotron emission \citep{2010ApJ...721L..82C,2014IJMPD..2330007B}.

\subsection{Properties of Faint Radio Halos}
The clusters ACT-CL J1137 and ACT-CL J0546 highlight the critical role of MeerKAT’s sensitivity combined with advanced imaging techniques (e.g., compact source subtraction and low-resolution mapping) in detecting faint diffuse radio emission at high redshifts. These systems exhibit lower radio powers, with $P_{1.4\,\mathrm{GHz}} = (0.30-0.53)\, \times 10^{25}\, \mathrm{W\,Hz^{-1}}$,  placing them at the faint end of the $P_{\mathrm{1.4\,GHz}}-M\mathrm{_{500c}^{unc}}$ scaling relation. Notably, these clusters also represent the lowest mass systems in our sample, consistent with the expectation that turbulent re-acceleration scales with cluster mass \citep{2010ApJ...721L..82C,2013ApJ...777..141C}. Higher mass mergers generate stronger turbulence within the ICM, amplifying magnetic fields and enhancing synchrotron radiation efficiency \citep{2014IJMPD..2330007B}. The faint emissions observed in these clusters reflect the reduced turbulence and weaker magnetic field amplification associated with their lower masses.
\newline
\newline
In ACT-CL J1137, faint diffuse emission emerges only after compact source subtraction, appearing centrally located and broadly aligned with the cluster's SZ peak. While its low power and limited extent place it at the faint end of the radio halo population, the marginal detection significance warrants cautious interpretation and further observational follow-up. ACT-CL J0546 presents a classification challenge. Although statistically significant diffuse emission is detected at the 3$\sigma$ level after compact source subtraction, the morphology is patchy and appears to trace individual galaxies rather than a centrally concentrated ICM component. Archival \textit{Chandra} X-ray data reveal a near-symmetric and relaxed ICM morphology, and no spatial correlation is observed between the radio emission and the X-ray structure (Figure \ref{fig:5}). These factors, combined with its optical morphology (right panel Figure \ref{fig:9}), suggest a near-relaxed dynamical state and limited turbulent activity.
\newline
\newline
Despite their faintness, these detections confirm that diffuse radio emission persists in the most massive, high-redshift clusters. The localisation of emission within $r_{500c}$ in these systems suggests a strong connection between the thermal and non-thermal components of the ICM. This connection is further supported by their alignment with the  $P_{\mathrm{1.4\,GHz}}-M\mathrm{_{500c}^{unc}}$ scaling relation, even at the low-power end.
\newline
\newline
These faint radio halos, detected only after careful compact source subtraction and low-resolution imaging, underscore the observational challenges of studying diffuse emission at high redshift. Our results highlight the importance of deeper and more sensitive observations to probe the faint end of the radio halo population and to advance our understanding of turbulence and magnetic field evolution in distant clusters.

\subsection{Scaling Relation and Implications}
\begin{figure}
 \includegraphics[width=0.48\textwidth]{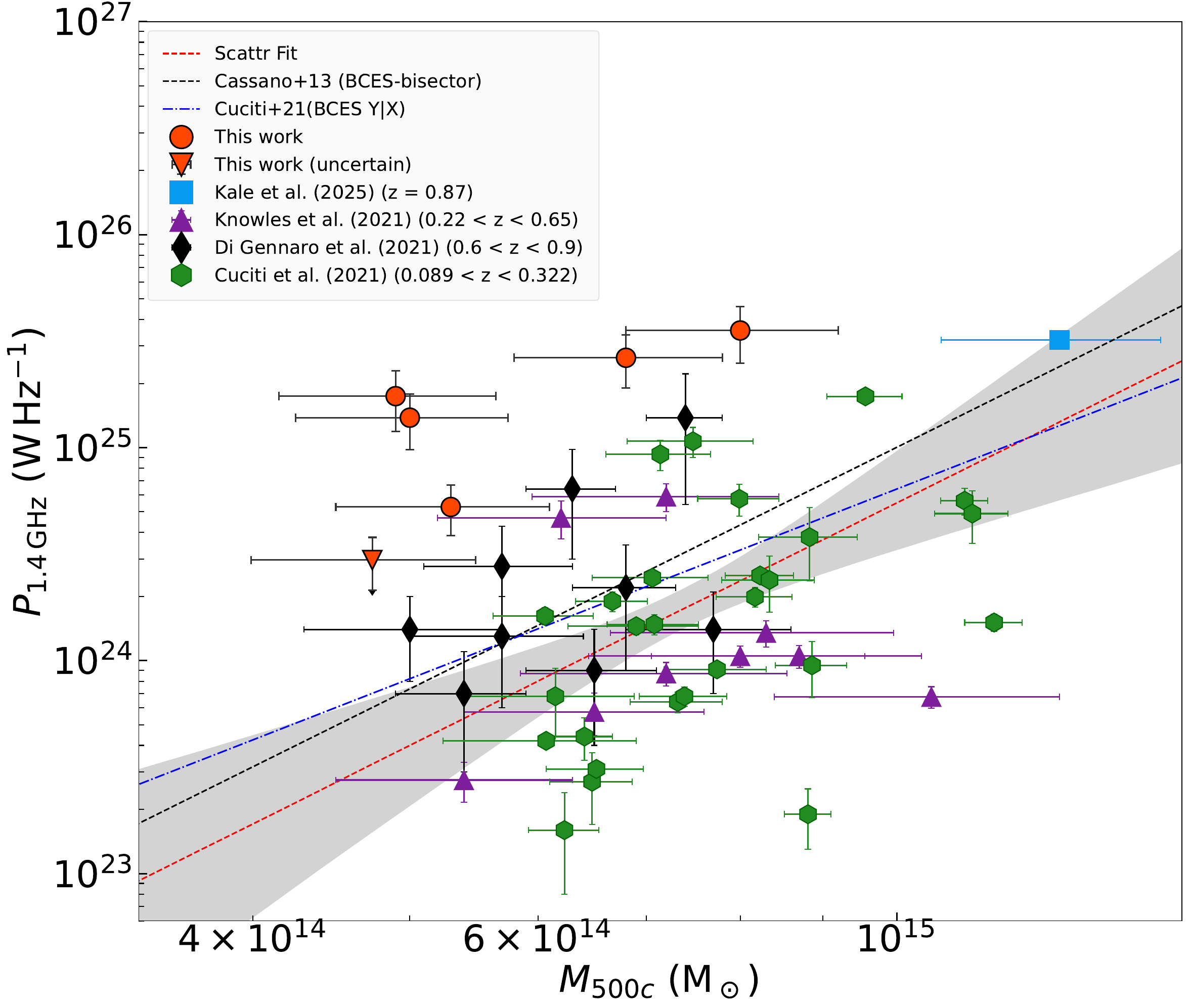}
 \caption{$P_{1.4\, \mathrm{GHz}}$--$M_{\mathrm{500c}}^{\mathrm{unc}}$ correlation plot for radio halos listed in Table \ref{table:3}. To address discrepancies between the mass estimates from ACT and Planck, we use the uncorrected $\mathrm{M_{500c}}$ values from ACT DR5 \citep{2021ApJS..253....3H}. The sample from \citet{2021NatAs...5..268D} includes radio halos at {\it z} $>$ 0.6, while the sample from \citet{2021MNRAS.504.1749K} spans 0.22 $<$ {\it z} $<$ 0.65.The green hexagonal points represent the sample from \citet{2021A&A...647A..51C} covering 0.089 $<$ {\it z} $<$  0.322. Additionally, the cluster El Gordo, located at {\it z} = 0.87, is incorporated into the analysis \citep{2025A&A...698A.271K}. The red dashed line and shaded regions represent the \texttt{scattr} best fit, with its 95$\%$ confidence interval. We also show the $P_{1.4\, \mathrm{GHz}}$--$M_{\mathrm{500}}$ correlations from \citet[blue dot-dashed line]{2021A&A...647A..51C} and \citet[black dashed line]{2013ApJ...777..141C}, scaled to 1.4 GHz. Note that the ACT SZ uncorrected masses plotted here are not corrected for Eddington bias, enabling direct comparison with Planck PSZ2 masses.}
 \label{fig:6}
\end{figure}

The $P_{\mathrm{1.4\,GHz}}-M\mathrm{_{500c}^{unc}}$ scaling relation is an important tool for probing the link between a galaxy cluster’s mass and its non-thermal radio emission. Our observations, represented by red points in Figure \ref{fig:6}, extend the existing data on this relationship to higher redshifts ({\it z} $>$ 1; see Table \ref{table:1} for redshift values) and demonstrate that these high-redshift clusters scatter around the established trend measured at  $z \sim 0.2$ \citep{2013ApJ...777..141C}. Notably, while Figure \ref{fig:6} does not explicitly show a strong redshift evolution, there appears to be a mild systematic offset. The data from \cite{2021MNRAS.504.1749K} mostly lie below the dashed line, whereas our data and those from \cite{2021NatAs...5..268D} lie above it. This pattern could suggest a mild redshift dependence, possibly reflecting evolving turbulence and magnetic field strengths in the ICM. If turbulent re-acceleration remains the dominant mechanism, this may imply stronger magnetic fields and turbulence levels at higher redshifts. Given that inverse Compton losses are higher at high redshift, maintaining radio halo luminosities comparable to low-$z$ systems requires either enhanced turbulent energy injection or magnetic field strengths higher than in local clusters, as predicted by turbulent re-acceleration models \citep{2021NatAs...5..268D}.
\newline
\newline
To quantify the relationship,  we performed a linear regression in log-log space using the \texttt{scattr}\footnote[16]{\url{https://github.com/lucadimascolo/scattr}} method, which accounts for intrinsic scatter and measurement uncertainties in both variables. The results of the fit are reported in Table \ref{table:4}, Our analysis combines our data with low-redshift samples from \citet{2021NatAs...5..268D} and \citet{2021MNRAS.504.1749K} . The results confirm that radio halo power increases steeply with cluster mass, consistent with previous findings. This steep scaling is often attributed to the increased turbulence present in massive, merging clusters, which efficiently accelerates relativistic particles and amplifies magnetic fields within the ICM, leading to enhanced synchrotron radiation.
\newline
\newline
Our findings are consistent with previous studies \citep{2010ApJ...721L..82C,2013ApJ...777..141C,2021A&A...647A..51C,2023A&A...680A..30C} that link mergers to the $P_{\mathrm{1.4\,GHz}}-M\mathrm{_{500c}^{unc}}$ scaling relation. As shown in Figure \ref{fig:6},  the clusters in our sample populate the same region of this plane as lower-redshift systems, consistent with turbulent re-acceleration as the dominant mechanism. However, while our data support this model, they do not definitively rule out alternative scenarios such as hadronic processes. Hadronic models, which posit that secondary electrons from proton-proton collisions produce synchrotron emission, predict a steeper dependence of radio power on cluster mass \citep{2005MNRAS.363.1173B,2014IJMPD..2330007B}. Furthermore, given our small sample size (six clusters), it remains challenging to disentangle mass and redshift effects definitively. Future studies with larger samples will be essential to better constrain these dependencies and explore potential deviations at {\it z} $>$ 1.

\subsection{SZ Mass Boosts in Merging Clusters}
The dynamical disturbance observed in 80$\%$ of our sample (4/5 clusters with X-ray data) has critical implications for SZ mass estimates. Hydrodynamical simulations demonstrate that mergers transiently enhance the integrated SZ signal (by $\sim 10-30\%$) due to shock-driven increases in thermal pressure, with these boosts persisting for $\sim 0.5-1\, \mathrm{Gyr}$ post-collision \citep{2008ApJ...679.1173R, 2008ApJ...680...17W}. At high redshifts ({\it z} $>$ 1), where mergers dominate hierarchical growth \citep{2007MNRAS.380..437P, 2025A&A...693A.106V}, such pressure enhancements may lead to inflated SZ-derived mass estimates compared to relaxed systems. Our sample, selected for high SZ masses ($M_{\rm 500c} > 4.5 \times 10^{14}$\,M$_{\odot}$), likely represents systems where merger-driven pressure boosts contribute to elevated $M\mathrm{_{500c}^{SZ}}$ values. 
\newline
\newline
Critically, diffuse radio emission is detected in all six clusters, with four hosting prominent radio halos and two showing faint detections (see Sections \ref{sec:Prominent Diffuse Emission} and \ref{sec:Faint Diffuse Emission Detected}). The observed radio halo powers ($P_{\rm{1.4\, GHz}} \sim 0.3-3.6 \times 10^{25}\, \mathrm{W\,Hz^{-1}}$) align with turbulent re-acceleration models \citep{2010ApJ...721L..82C}, providing empirical evidence that the same mergers that may drive transient SZ signal boosts also inject sufficient turbulence to sustain radio halos. However, the faint detections (ACT-CL J1137.8+0728 and J0546.6–5345) require further investigation. Notably, ACT-CL J0546 exhibits a relaxed X-ray morphology but shows faint, patchy radio emission that may originate from AGN activity or fossil plasma from a minor merger rather than merger-driven turbulence \citep[e.g.,][]{2019A&A...622A..24S,2022A&A...661A..92B}. This suggests that while mergers dominate the sample, secondary processes may contribute to non-thermal emission in some high-$z$ clusters. The synergy between SZ and radio observations reinforces simulations showing that merger boosts preferentially select massive, dynamically active clusters in high-$z$ SZ catalogues \citep{2008ApJ...680...17W, 2012MNRAS.419.1766K}.

\section{SUMMARY AND CONCLUSIONS}
\label{sec:SUMMARY AND CONCLUSIONS}
In this study, we presented the results of a search for diffuse radio emission in six massive galaxy clusters at redshifts ${\it z} > 1$, selected from the MeerKAT Massive Distant Cluster Survey (MMDCS). These six clusters represent the most massive systems in the sample, as determined by their SZ mass, and the full MMDCS sample will be analysed in future work. Using advanced imaging techniques, including compact source subtraction and low-resolution mapping, we detected diffuse radio emission in four clusters (ACT-CL J2106, J1142, J0329, and J0003) and identified faint emission in two additional systems (ACT-CL J1137 and J0546). The radio halos exhibit a range of linear sizes from 0.47 Mpc to 1.08 Mpc and radio powers spanning $P_{1.4\,\mathrm{GHz}} = (0.30\, \pm\, 0.08) \times 10^{25}\, \mathrm{W\,Hz^{-1}}$ to $(3.55\, \pm\, 1.06) \times 10^{25}\, \mathrm{W\,Hz^{-1}}$. These results extend the known population of radio halos into the high-redshift regime, demonstrating that diffuse synchrotron emission persists in massive clusters even under the challenging conditions imposed by increased inverse Compton losses at {\it z} $>$ 1. 
\newline
The dynamical states of the clusters were investigated using Chandra X-ray imaging combined with MeerKAT radio observations. Four clusters (ACT-CL J2106, J1142, J0329, and J0003) display disturbed X-ray morphologies characterised by asymmetries, elongation, and substructure, which are features indicative of ongoing or recent merger activity. These systems host prominent radio halos tracing turbulent regions within $R_{\rm 500c}$, reinforcing the connection between mergers and non-thermal processes in the ICM. In contrast, ACT-CL J0546 exhibits a relaxed morphology with no clear halo signature, while ACT-CL J1137 remains unknown due to the lack of X-ray data.
\newline
\newline
The scaling relation between radio halo power and cluster mass ($P_{\mathrm{1.4\,GHz}}-M\mathrm{_{500c}^{unc}}$) was analysed, revealing that our high-redshift sample aligns with trends observed at lower redshifts, supporting turbulent re-acceleration as the dominant mechanism for radio halo formation. While no explicit redshift-dependent trend was observed, our data points are consistent with the low-redshift, high-mass region of the scaling relation. This suggests mergers in massive clusters inject sufficient turbulence to sustain luminous halos despite enhanced energy losses due to the CMB. The faint radio emission in ACT-CL J0546 might reflect reduced turbulent activity in relaxed systems, consistent with models where merger-driven shocks dominate particle acceleration. However, the scarcity of confirmed relaxed clusters at $z > 1$ currently limits conclusions about this relationship. The consistency of the $P_{\mathrm{1.4\,GHz}}-M\mathrm{_{500c}^{unc}}$ scaling relation across redshifts supports turbulent re-acceleration as the dominant mechanism for radio halo formation in merging clusters. Mergers are essential for generating the turbulence required to power radio halos, as evidenced by the disturbed dynamical states of the majority of our sample. While hydrodynamical simulations suggest merger-driven thermal pressure boosts can temporarily elevate SZ masses by $\sim 10-30\%$ \citep{2008ApJ...680...17W, 2012MNRAS.419.1766K}, this transient effect does not uniformly bias SZ-derived masses. Relaxed clusters, though less common in high-z SZ surveys, may exhibit brighter radio emission from AGN or cool-core-related processes, highlighting the complex interplay between dynamical state and non-thermal emission mechanisms.
\newline
\newline
In conclusion, our findings demonstrate MeerKAT’s unique capability to detect faint diffuse emission in high-redshift clusters, providing critical insights into the interplay between cluster mass, dynamical state, and non-thermal processes in the early universe. However, spectral index measurements remain challenging due to current sensitivity limits. Diffuse radio emission is detected in all six clusters (4 prominent halos, 2 faint), underscoring the prevalence of merger-driven turbulence in massive $z > 1$ systems. Future observations at lower frequencies (e.g., MeerKAT’s $UHF$-band) will be essential for constraining magnetic field strengths and electron energy loss mechanisms through detailed spectral studies. Expanding the analysis to the full MMDCS sample of 30 clusters will clarify whether the high detection rate persists across a broader mass range or reflects a unique property of the most extreme $z > 1$ mergers. This will enable a comprehensive investigation into the evolution and growth of magnetic fields across cosmic time, shedding light on how these fields develop in massive galaxy clusters. 

\section*{Acknowledgements}
The MeerKAT telescope is operated by the South African Radio Astronomy Observatory, which is a facility of the National Research Foundation, an agency of the Department of Science and Innovation. The financial assistance of the SARAO towards this
research is hereby acknowledged. MH, KM, and US acknowledge support from the National Research Foundation of South Africa.

\section*{Data Availability}

The raw data underlying this article are available in SARAO archive under the proposal ID SCI-20220822-MH-01, at \text{\url{https://archive.sarao.ac.za}}.


\bibliographystyle{mnras}
\bibliography{reference} 



\appendix
\makeatletter
\def\@capstart{} 
\def\uppercaseDECaLS{DECaLS}
\renewcommand{\MakeUppercase}[1]{%
  \def\temp{#1}%
  \ifx\temp\uppercaseDECaLS
    DECaLS%
  \else
    \MakeTextUppercase{#1}%
  \fi
}
\makeatother

\section{DEC\MakeLowercase{a}LS optical images}

\begin{figure*}
\begin{multicols}{2}
\includegraphics[width=0.48\textwidth]{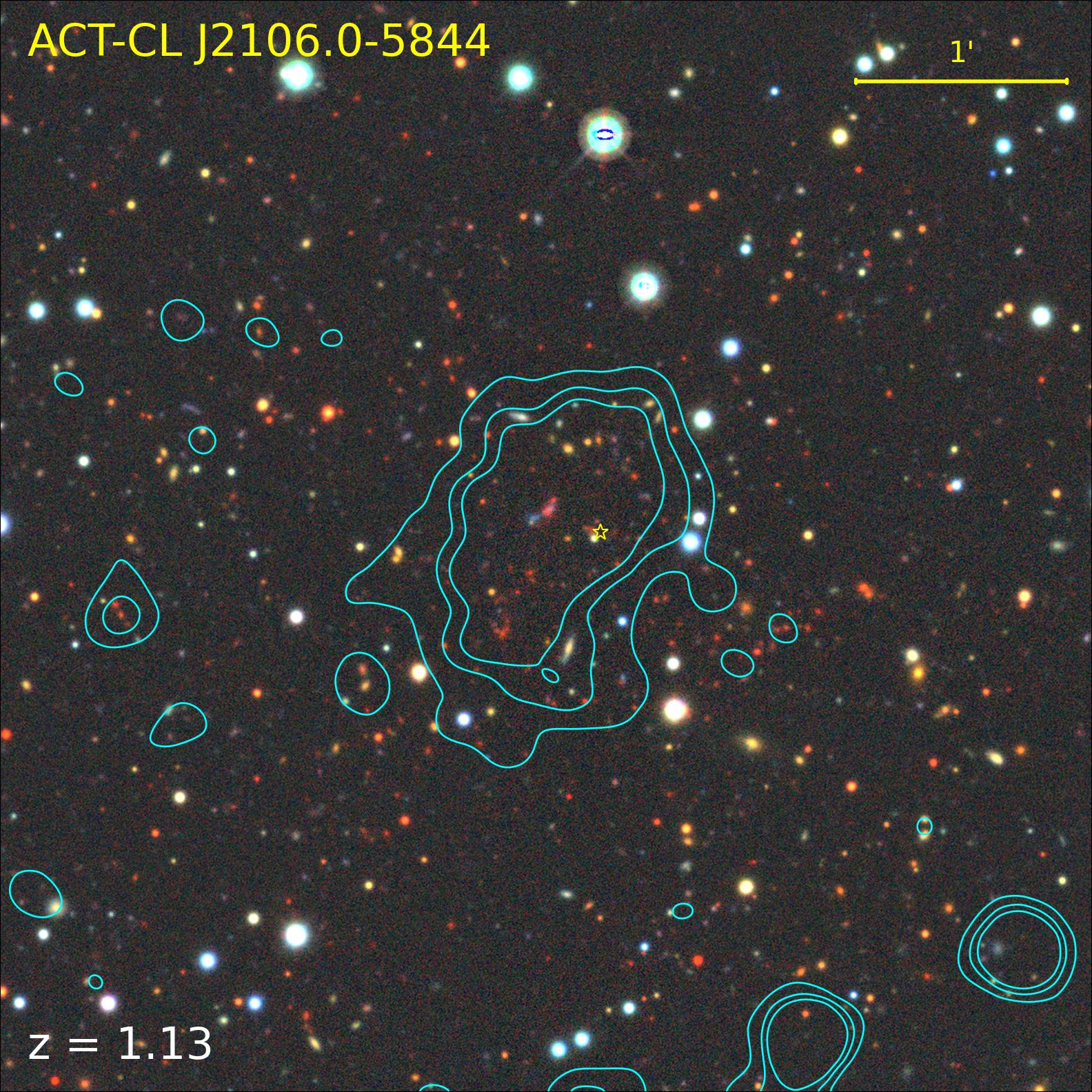}\par 
\includegraphics[width=0.48\textwidth]{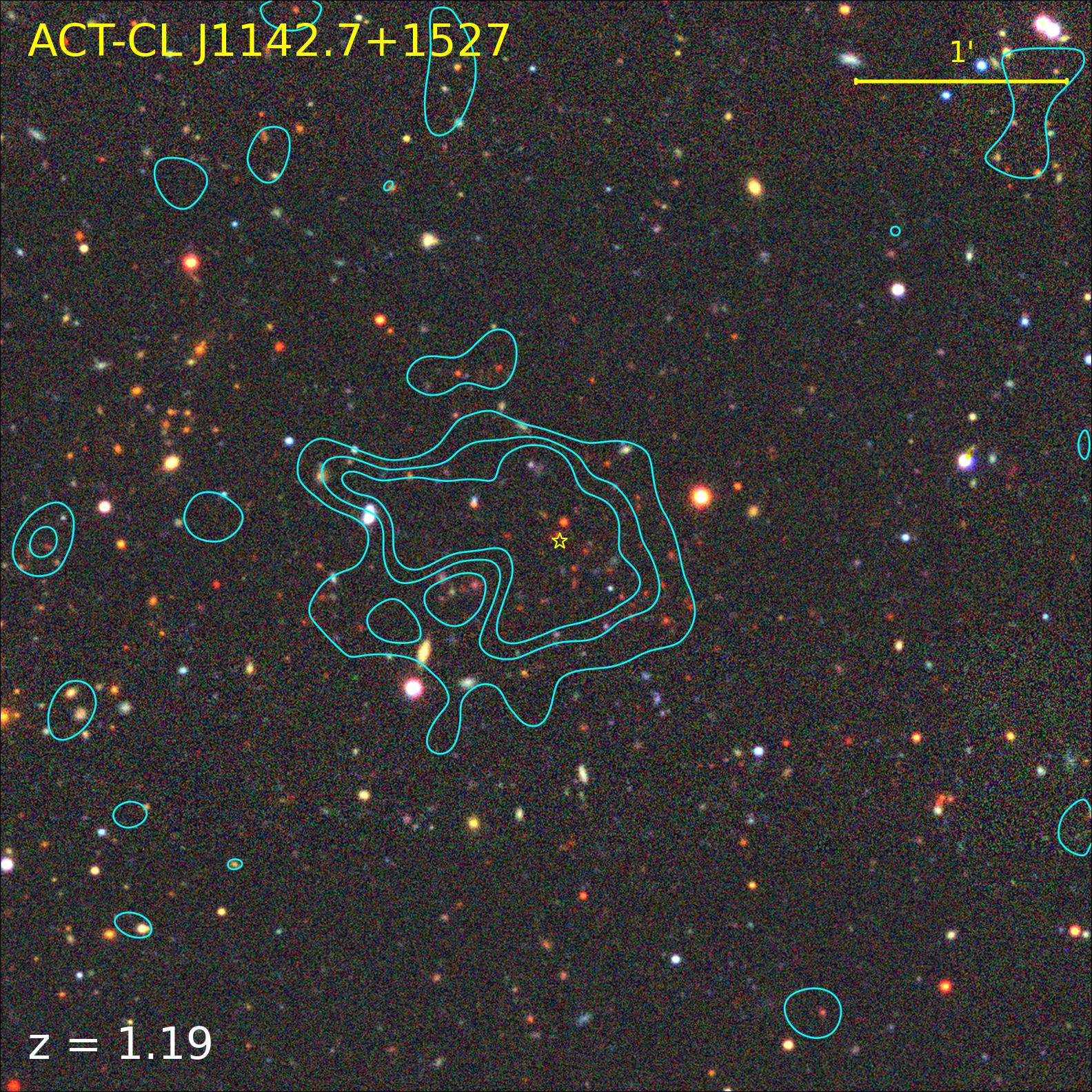}\par 
\end{multicols}
\caption{The Optical \textit{grz} images obtained from the DECaLS survey \citep{2019AJ....157..168D} are shown with MeerKAT $L$-band LR radio contours overlaid in cyan at levels of  $\mathrm{\sigma_{LR} \times [3, 6, 10]}$. The yellow star indicates the position of the BCG, adopted from the ACT DR5 catalogue \citep{2021ApJS..253....3H}. \textit{Left}: ACT-CL J2106, with a $1\sigma$ noise level of $6.5\, \mathrm{\mu Jy/beam}$. \textit{Right}: ACT-CL J1142, with a $1\sigma$ noise level of $7.2\, \mathrm{\mu Jy/beam}$.}
\label{fig:7}
\end{figure*}

\begin{figure*}
\begin{multicols}{2}
\includegraphics[width=0.48\textwidth]{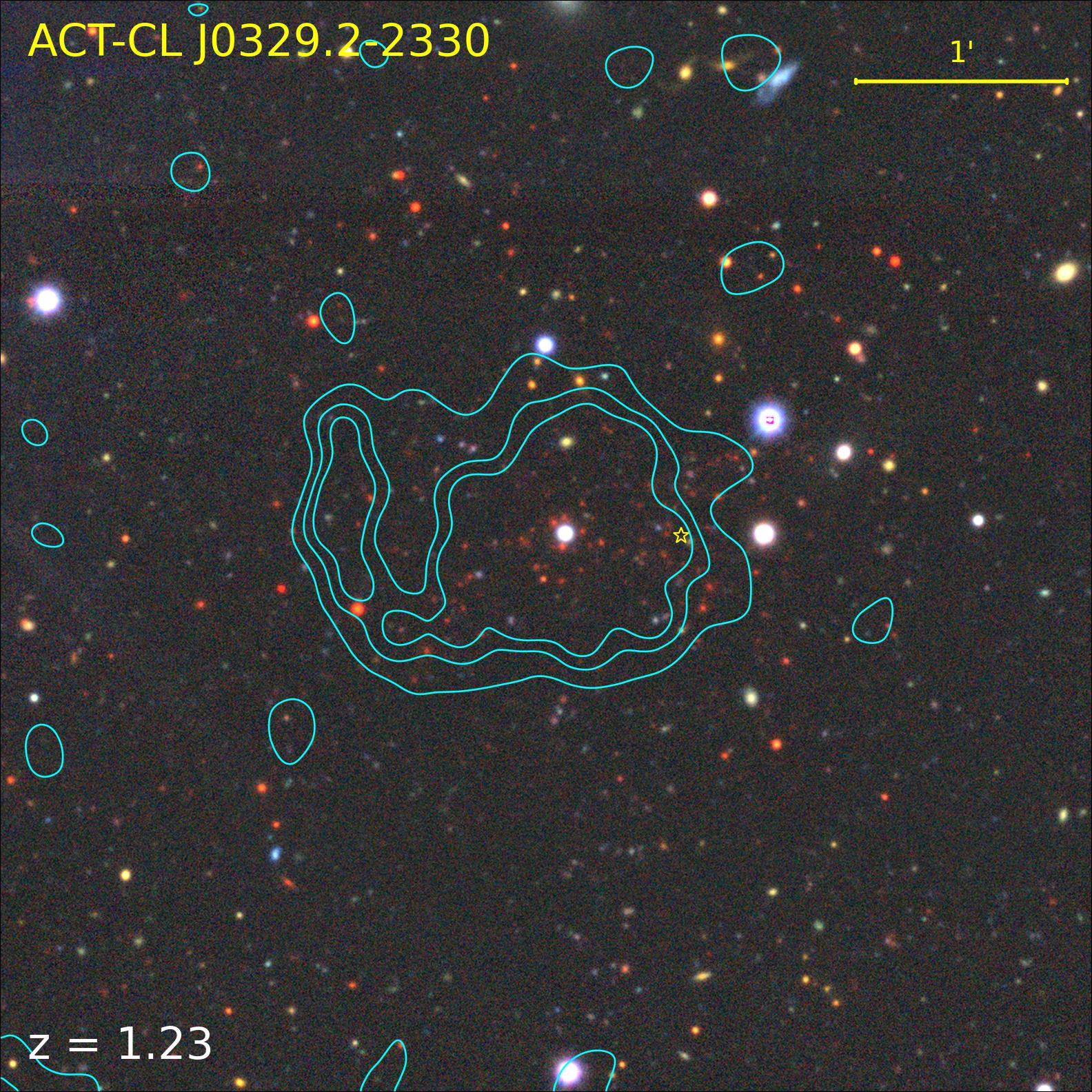}\par 
\includegraphics[width=0.48\textwidth]{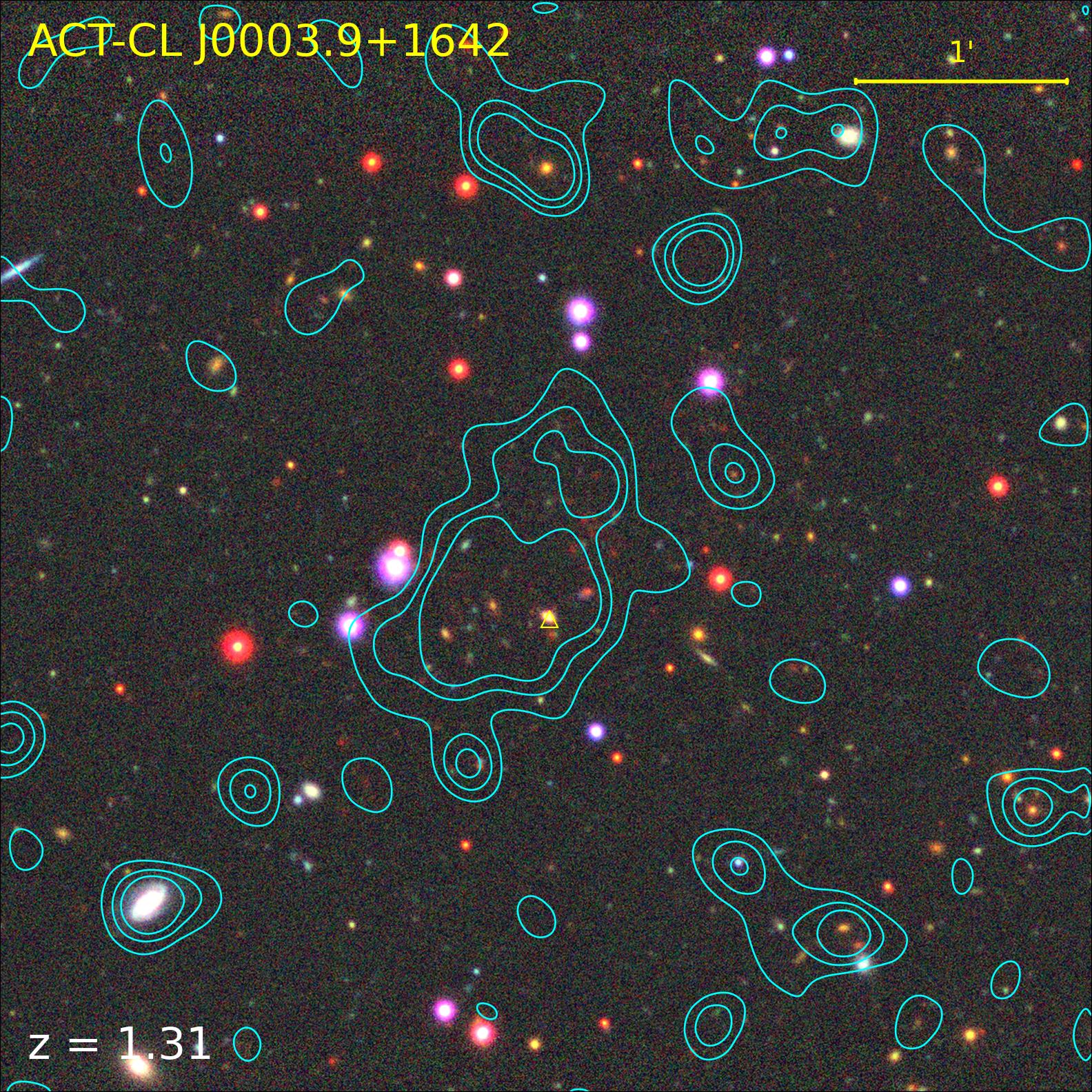}\par 
\end{multicols}
\caption{The Optical \textit{grz} images obtained from the DECaLS survey \citep{2019AJ....157..168D} are shown with MeerKAT $L$-band LR radio contours overlaid in cyan at levels of  $\mathrm{\sigma_{LR} \times [3, 6, 10]}$. The yellow star/triangle indicates the position of the BCG, adopted from the ACT DR5 catalogue \citep{2021ApJS..253....3H}. \textit{Left}: ACT-CL J0329, with a $1\sigma$ noise level of $6.1\, \mathrm{\mu Jy/beam}$. \textit{Right}: ACT-CL J0003, with a $1\sigma$ noise level of $9.2\, \mathrm{\mu Jy/beam}$.}
\label{fig:8}
\end{figure*}

\begin{figure*}
\begin{multicols}{2}
\includegraphics[width=0.48\textwidth]{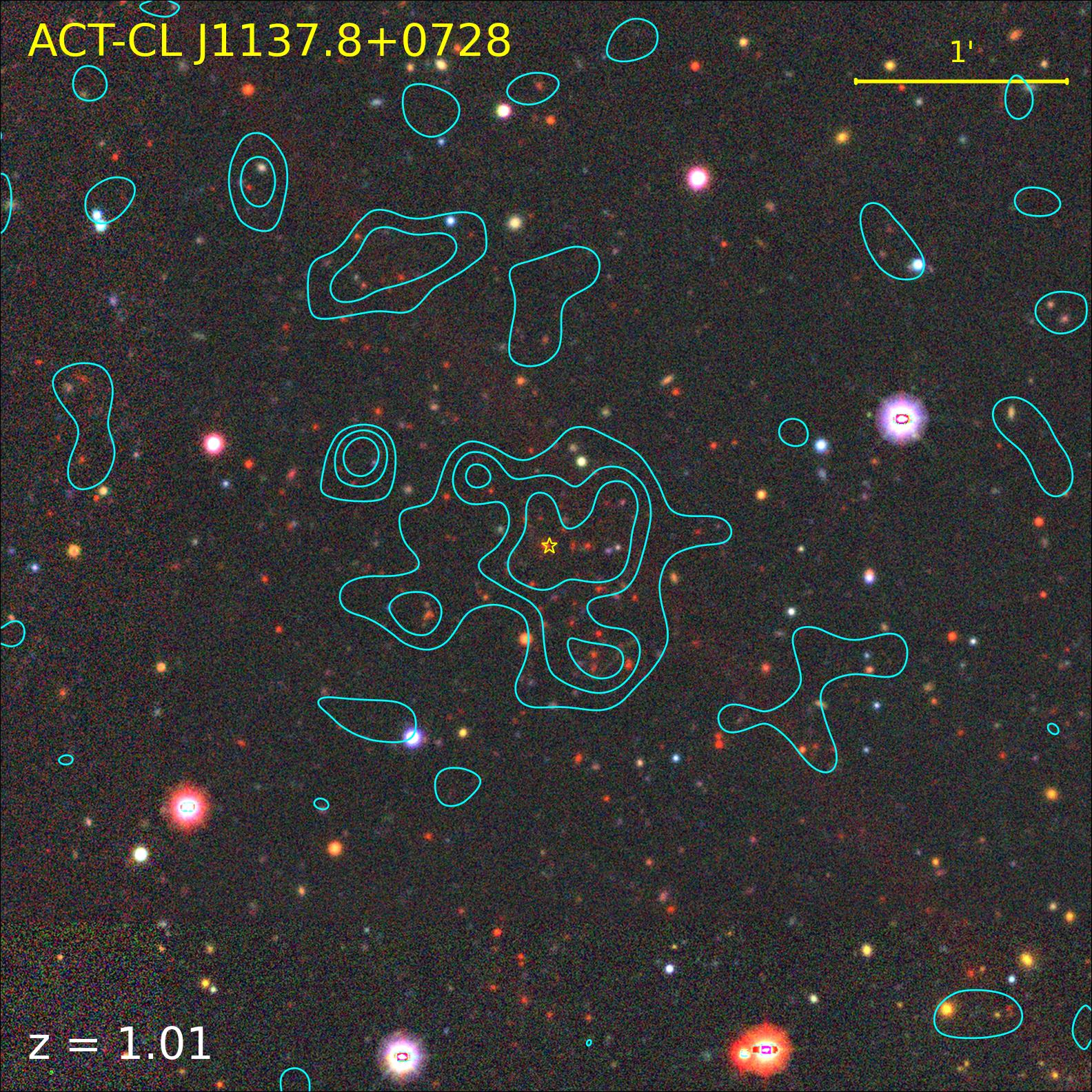}\par 
\includegraphics[width=0.48\textwidth]{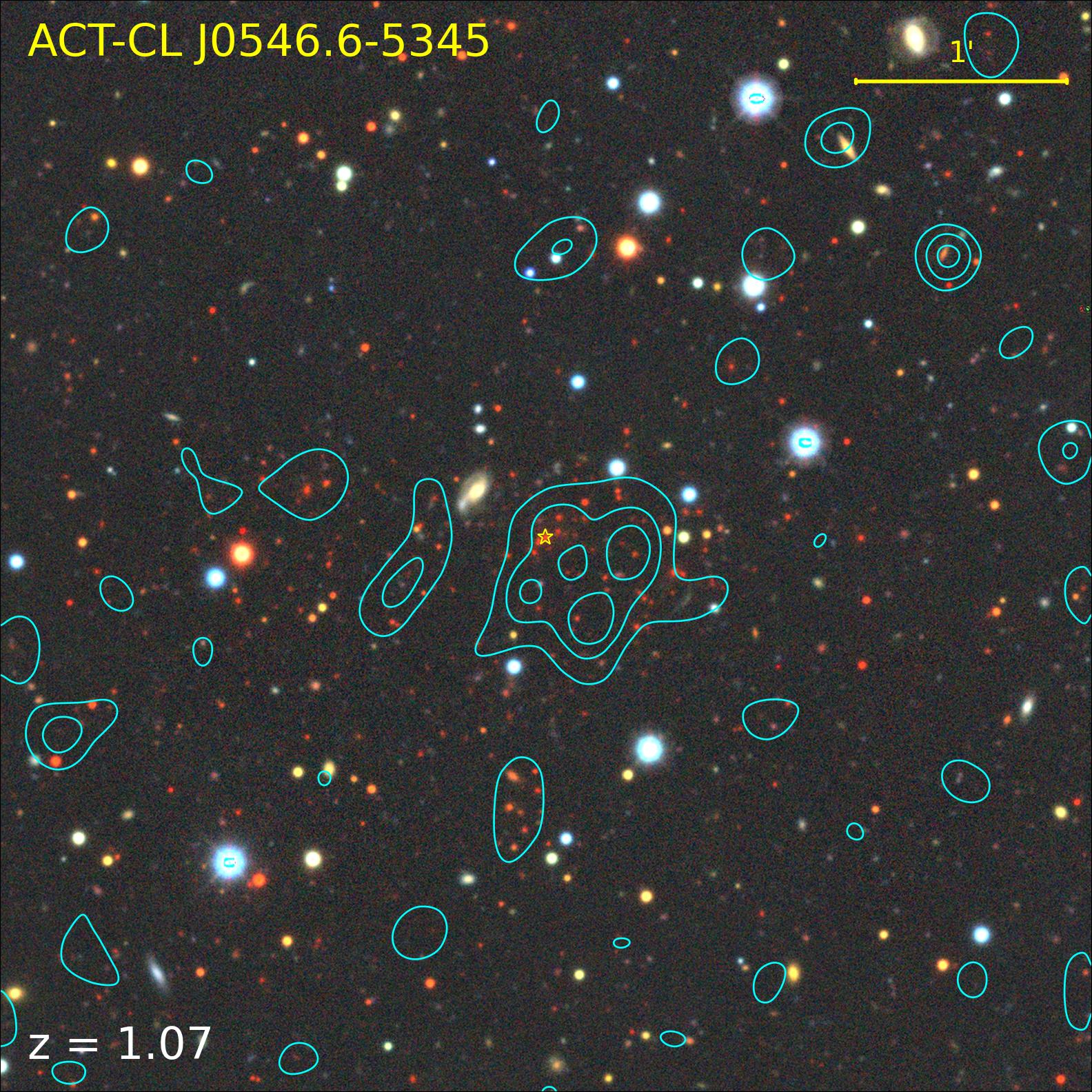}\par 
\end{multicols}
\caption{The Optical \textit{grz} images obtained from the DECaLS survey \citep{2019AJ....157..168D} are shown with MeerKAT $L$-band LR radio contours overlaid in cyan at levels of $\mathrm{\sigma_{LR} \times [3, 6, 10]}$. The yellow star indicates the position of the BCG, adopted from the ACT DR5 catalogue \citep{2021ApJS..253....3H}. \textit{Left}: ACT-CL J1137, with a $1\sigma$ noise level of $7.5\, \mathrm{\mu Jy/beam}$. \textit{Right}: ACT-CL J0546, with a $1\sigma$ noise level of $9.4\, \mathrm{\mu Jy/beam}$.}
\label{fig:9}
\end{figure*}
\section{High and low resolution images}

\begin{figure*}
    \centering
    \includegraphics[width=\linewidth]{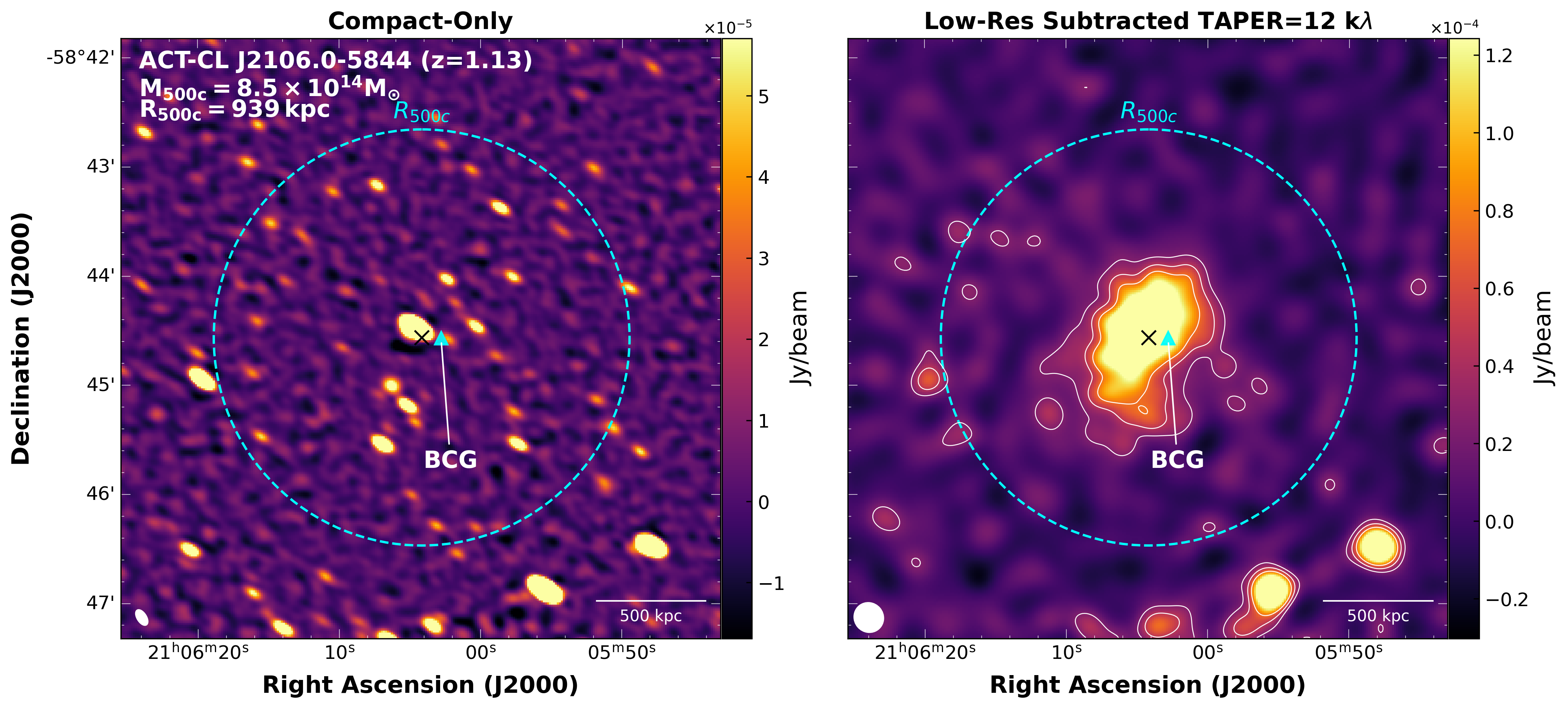}
    \caption{MeerKAT $L$-band radio images of ACT-CL J2106.0-5844. The position of the BCG is indicated by a cyan triangle, while the black cross marks the location of the ACT SZ peak. \textit{Left}: Compact-only image showing discrete radio sources. \textit{Right}: Low-resolution, point-source subtracted image with white contours at levels of $\sigma_{\mathrm{LR}} \times [3,6,10]$; the $1\sigma$ noise level is $9.4\, \mathrm{\mu Jy/beam}$.
    }
    \label{fig:10}
\end{figure*}

\begin{figure*}
    \centering
    \includegraphics[width=\linewidth]{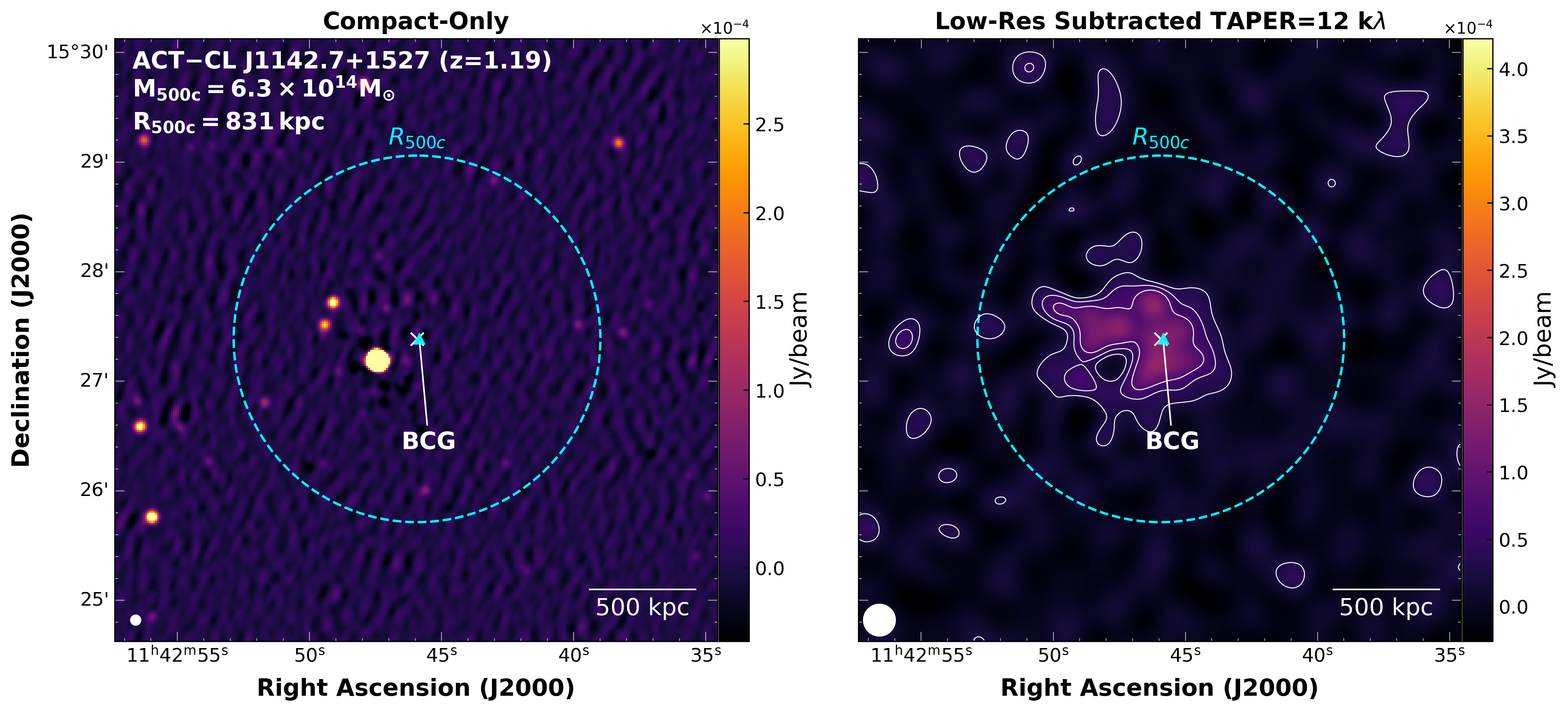}
    \caption{MeerKAT $L$-band radio images of ACT-CL J1142.7+1527. The position of the BCG is indicated by a cyan triangle, while the black cross marks the location of the ACT SZ peak. \textit{Left}: Compact-only image showing discrete radio sources. \textit{Right}: Low-resolution, point-source subtracted image with white contours at levels of $\sigma_{\mathrm{LR}} \times [3,6,10]$; the $1\sigma$ noise level is $6.5\, \mathrm{\mu Jy/beam}$.
    }
    \label{fig:11}
\end{figure*}

\begin{figure*}
    \centering
    \includegraphics[width=\linewidth]{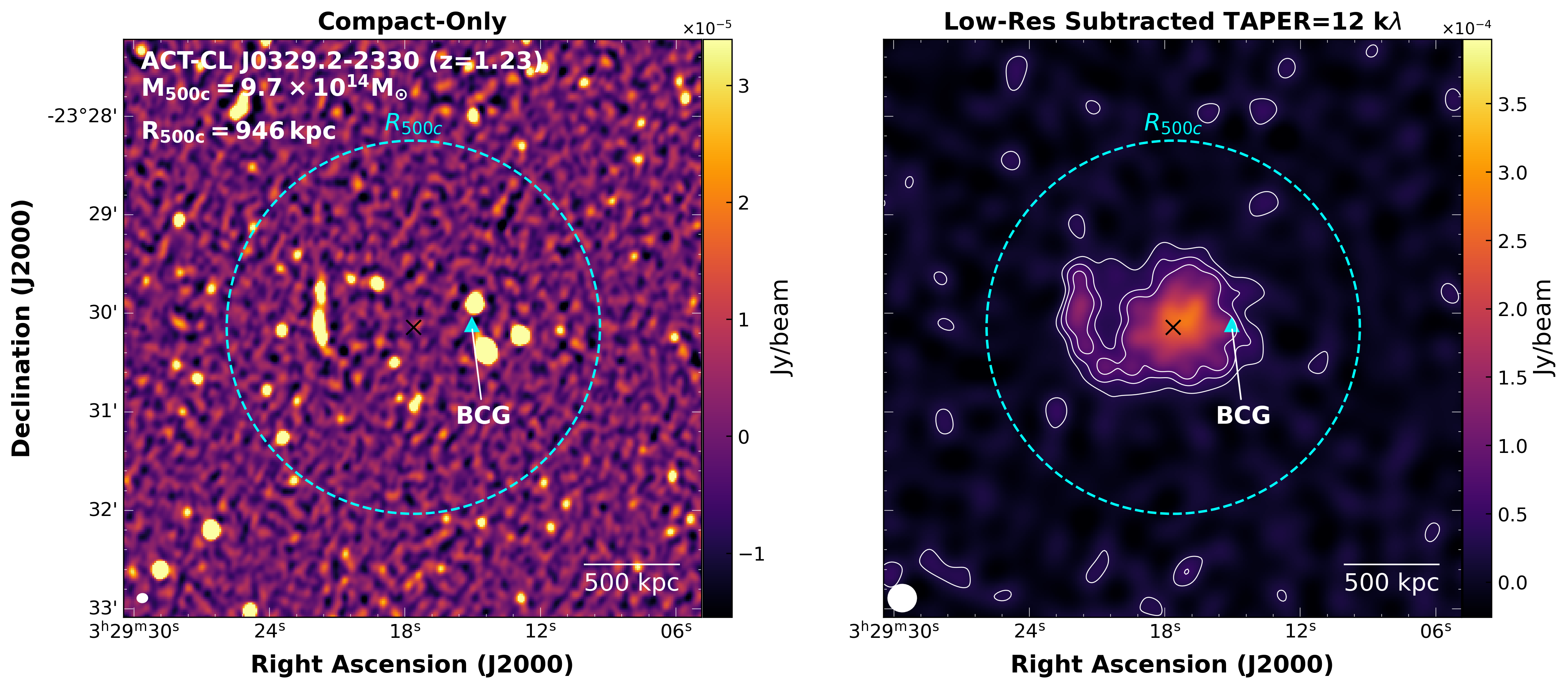}
    \caption{MeerKAT $L$-band radio images of ACT-CL J0329.2-2330. The position of the BCG is indicated by a cyan triangle, while the black cross marks the location of the ACT SZ peak. \textit{Left}: Compact-only image showing discrete radio sources. \textit{Right}: Low-resolution, point-source subtracted image with white contours at levels of $\sigma_{\mathrm{LR}} \times [3,6,10]$; the $1\sigma$ noise level is $6.1\, \mathrm{\mu Jy/beam}$.
    }
    \label{fig:12}
\end{figure*}

\begin{figure*}
    \centering
    \includegraphics[width=\linewidth]{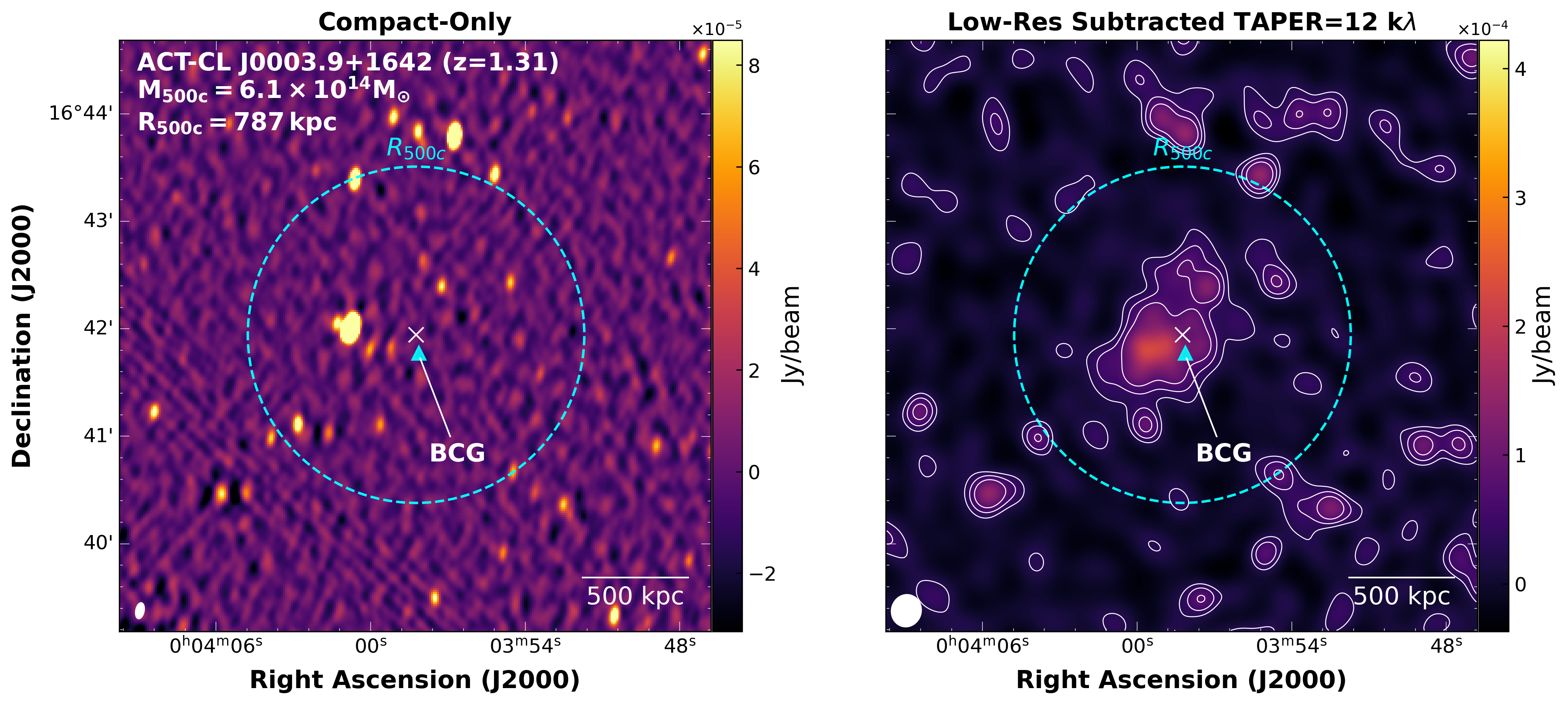}
    \caption{MeerKAT $L$-band radio images of ACT-CL J0003.9+1642. The position of the BCG is indicated by a cyan triangle, while the black cross marks the location of the ACT SZ peak. \textit{Left}: Compact-only image showing discrete radio sources. \textit{Right}: Low-resolution, point-source subtracted image with white contours at levels of $\sigma_{\mathrm{LR}} \times [3,6,10]$; the $1\sigma$ noise level is $9.2\, \mathrm{\mu Jy/beam}$.
    }
    \label{fig:13}
\end{figure*}

\begin{figure*}
    \centering
    \includegraphics[width=\linewidth]{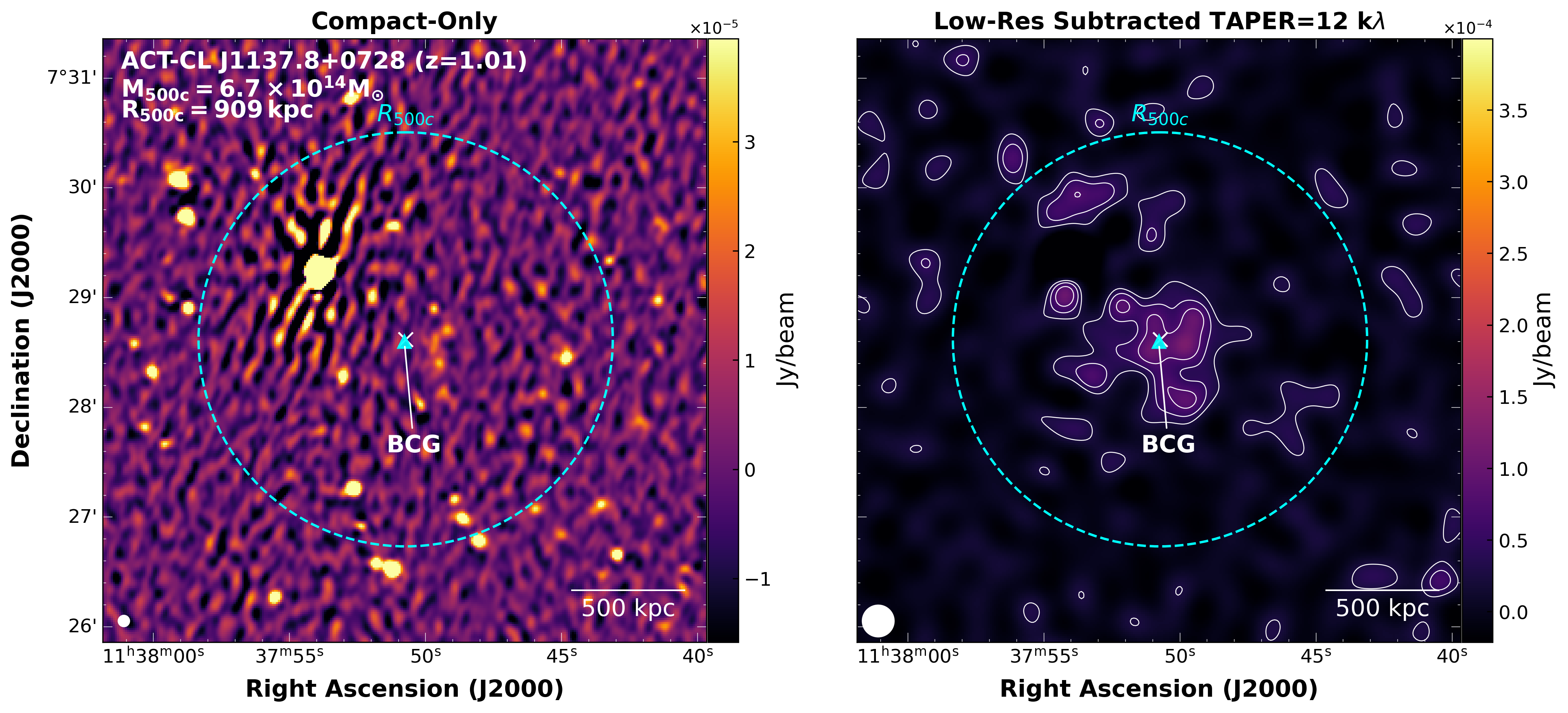}
    \caption{MeerKAT $L$-band radio images of ACT-CL J1137.8+0728. The position of the BCG is indicated by a cyan triangle, while the black cross marks the location of the ACT SZ peak. \textit{Left}: Compact-only image showing discrete radio sources. \textit{Right}: Low-resolution, point-source subtracted image with white contours at levels of $\sigma_{\mathrm{LR}} \times [3,6,10]$; the $1\sigma$ noise level is $7.5\, \mathrm{\mu Jy/beam}$.
    }
    \label{fig:14}
\end{figure*}

\begin{figure*}
    \centering
    \includegraphics[width=\linewidth]{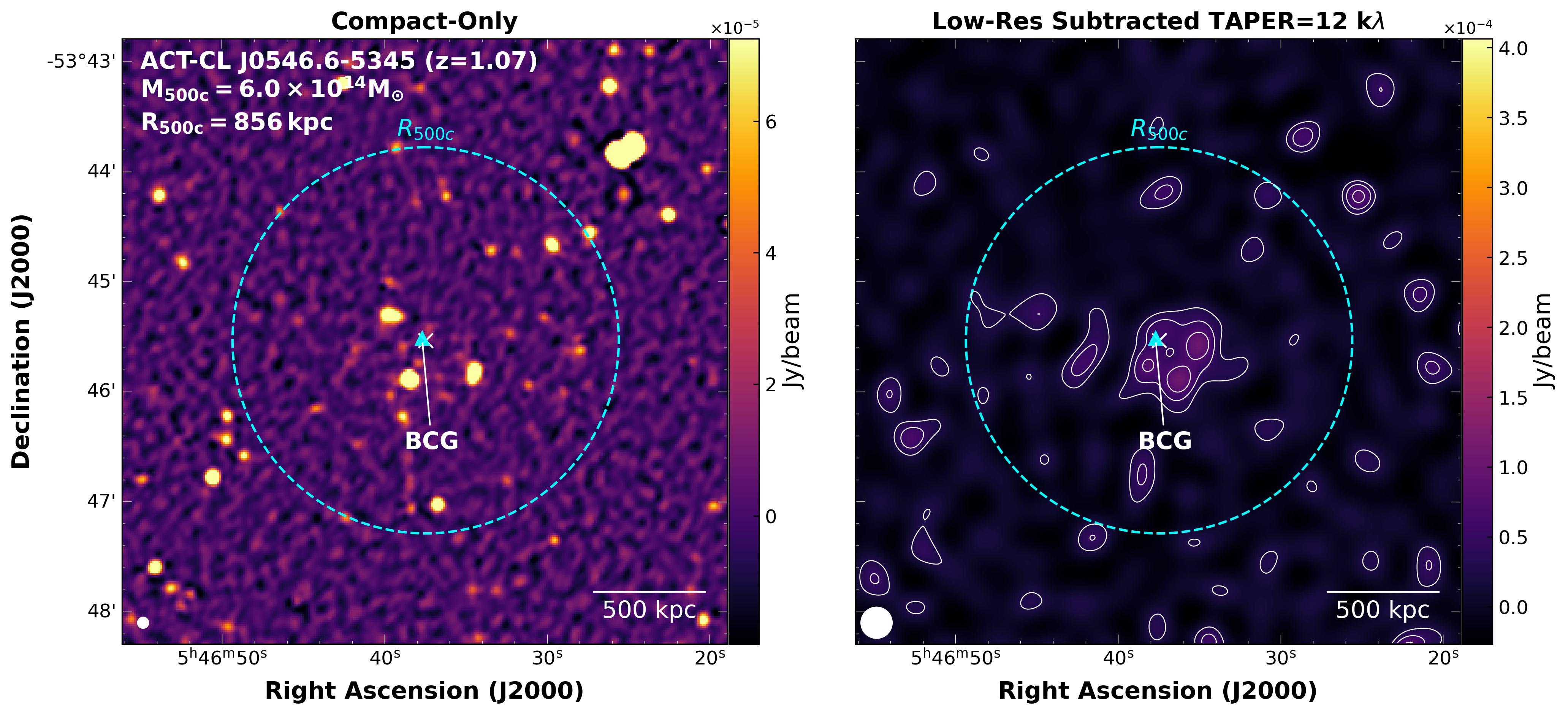}
    \caption{MeerKAT $L$-band radio images of ACT-CL J0546.6-5345. The position of the BCG is indicated by a cyan triangle, while the black cross marks the location of the ACT SZ peak. \textit{Left}: Compact-only image showing discrete radio sources. \textit{Right}: Low-resolution, point-source subtracted image with white contours at levels of $\sigma_{\mathrm{LR}} \times [3,6,10]$; the $1\sigma$ noise level is $9.4\, \mathrm{\mu Jy/beam}$. }
    \label{fig:15}
\end{figure*}

\section{Fitting procedure and parameters}
\begin{table}
\caption{ Fitting parameters from \texttt{scattr}, along with the parameters of the scaling relation from \citet{2021A&A...647A..51C} and \citet{2013ApJ...777..141C}.}
\label{table:4}
\begin{tabular}{ccccc}
\hline
Method        & \textit{B} & err(\textit{B}) & \textit{A} & err(\textit{A})  \\ \hline
\texttt{scattr}        & 3.83      & 1.52            & -0.01       & 0.10            \\
BCES Y|X      & 2.96       & 0.50            & 0.01       & 0.10            \\
BCES bisector & 3.77       & 0.57            & 0.125      & 0.076           \\ \hline
\end{tabular}

\end{table}

We estimated the correlation parameters for radio halos by fitting a power-law model to the $P_{1.4\, \mathrm{GHz}}$--$M_{\mathrm{500}}$ relation in log-log space,following the approach of \citet{2013ApJ...777..141C}. The fitting was performed using \texttt{scattr}, a regression tool that handles intrinsic scatter and uncertainties in both variables. The fitted relation is expressed as
\begin{equation}
  \log\left( \frac{P_{1.4\,\text{GHz}}}{10^{24.5}\,\text{W\,Hz}^{-1}} \right) = B \log\left( \frac{M_{500}}{10^{14.9}\,M_\odot} \right) + A,  
\end{equation}
where $A$ is the intercept and $B$ the slope. Our results give $B\, =\, 3.83\, \pm\, 1.52$ and  $A\, =\, -0.01\, \pm\, 0.10$, which are comparable to those derived using BCES methods in previous studies (see Table \ref{table:4}).


\bsp	
\label{lastpage}
\end{document}